\documentclass[3p,review]{elsarticle}
\usepackage{bm}
\usepackage{graphicx}
\usepackage{color}
\usepackage{subfigure}
\usepackage{hyperref}
\usepackage{latexsym}
\usepackage{amsthm}
\usepackage{amssymb}
\usepackage{amsmath}
\usepackage{wrapfig}
\usepackage{ulem}
\biboptions{sort&compress}
\usepackage{amsmath,amsfonts,amssymb,bm,cancel}
\usepackage{soul,cancel}
\usepackage{epsf,amsmath,amssymb,verbatim,color,multirow,pifont,graphicx,cleveref,tabularx}

\usepackage{amssymb}
\usepackage{amsmath}
\usepackage{amsthm}
\usepackage{mathtools}
\usepackage{mathrsfs}

\usepackage[figuresright]{rotating}

\begin{document}

\theoremstyle{plain}
\newtheorem{lemma}{Lemma}
\def\be{\begin{equation}}
\def\ee{\end{equation}}

\def\bc{\begin{center}}
\def\ec{\end{center}}
\def\bea{\begin{eqnarray}}
\def\eea{\end{eqnarray}}
\newcommand{\avg}[1]{\langle{#1}\rangle}
\newcommand{\Avg}[1]{\left\langle{#1}\right\rangle}

\def\ie{\textit{i.e.}}
\def\etal{\textit{et al.}}
\def\m{\vec{m}}
\def\G{\mathcal{G}}
\begin{frontmatter}




\title{ Topology and dynamics of higher-order multiplex networks }


\author{Sanjukta Krishnagopal}

\address{Department of Electrical Engineering and Computer Science, University of California Berkeley, Berkeley, CA 94720, USA\\Department of Mathematics, University of California Los Angeles, Los Angeles, CA 90095, USA}
\author{Ginestra Bianconi}

\address{School of Mathematical Sciences, Queen Mary University of London, London, E1 4NS, United Kingdom\\
The  Alan  Turing  Institute,  96  Euston  Road,  London,  NW1  2DB,  United  Kingdom}
\begin{abstract}
Higher-order networks are gaining significant scientific attention due to their ability to encode the many-body interactions present in complex systems. However, higher-order networks have the limitation that they only capture many-body interactions of the same type.  To address this limitation, we present a mathematical framework that determines the topology of higher-order multiplex networks and illustrates the interplay between their topology and dynamics. Specifically, we examine the diffusion of topological signals associated not only to the nodes but also to the links and to the higher-dimensional simplices of multiplex simplicial complexes.
We leverage on the ubiquitous presence of the overlap of the simplices to couple the dynamics among multiplex layers, introducing a definition of multiplex Hodge Laplacians and Dirac operators. We show that the spectral properties of these operators determine higher-order diffusion on higher-order multiplex networks and encode their multiplex Betti numbers. Our numerical investigation of the spectral properties of synthetic and real (connectome, microbiome) multiplex simplicial complexes indicates that the coupling between the layers can either speed up or slow down the higher-order diffusion of topological signals. This mathematical framework is very general and can be applied to study generic higher-order systems with interactions of multiple types. In particular, these results might find applications in brain networks which are understood to be both multilayer and higher-order.  
\end{abstract}


\end{frontmatter}



\section{Introduction}

Higher-order networks~\cite{bianconi2021higher,battiston2020networks,bick2021higher,boccaletti2023structure,otter2017roadmap}, and in particular simplicial complexes, capture the many-body interactions present in complex systems and display a rich interplay between  topology and  dynamics \cite{battiston2021physics,majhi2022dynamics}. 
Important and unexpected higher-order collective phenomena having a distinct topological nature  have been shown to lead to  synchronization~\cite{millan2020explosive,ghorbanchian2021higher,carletti2023global,arnaudon2022connecting,deville2021consensus,calmon2023local,calmon2022dirac,mulas2020coupled,skardal2019abrupt,nurisso2023unified} and diffusion~\cite{torres2020simplicial,ziegler2022balanced,muhammad2006control,schaub2020random,krishnagopal2021spectral}  of topological signals, i.e. dynamical variables associated not only to nodes, but also to links, triangles, and higher dimensional simplices of simplicial complexes. Higher-order diffusion  as well as higher-order synchronization are dictated  by algebraic topological operators including the Hodge Laplacians \cite{eckmann1944harmonische,horak2013spectra,jost2019hypergraph,lim2020hodge,meng2020weighted,mulas2022graphs} and the Dirac operators \cite{bianconi2021topological,baccini2022weighted,wee2023persistent,bianconi2023dirac,bianconi2023mass} and involve the irrotational and  the solenoidal components of the topological signals, revealing a deep relation of these dynamical processes with the topology of the simplicial complexes. The concepts and ideas arising in this domain are complementary to recent research activity  \cite{reimann2017cliques,petri2014homological,santoro2023higher,faskowitz2022edges,santos2019topological}, pointing out the important role of topology in higher-order brain dynamics. Moreover, other dynamical processes determined by the topology of higher-order networks include higher-order Turing patterns~\cite{giambagli2022diffusion,muolo2023three}, epidemic spreading \cite{taylor2015topological}, as well as  percolation \cite{bobrowski2020homological,santos2019topological,sun2023dynamic}. Finally, topology has been shown to be key in machine learning and artificial intelligence, including  topological machine learning \cite{hensel2021survey},  topological data analysis (TDA) \cite{petri2014homological,petri2013topological,otter2017roadmap}  ranking~\cite{jiang2011statistical} and signal processing algorithms~\cite{barbarossa2020topological,schaub2021signal,calmon2023dirac}. 
While other works adopt a node-centered approach where dynamical variables are associated only to the nodes of simplicial complexes \cite{iacopini2019simplicial,skardal2020higher,jalan2022multiple,anwar2022stability,sun2022diffusion,fan2022epidemics,de2021phase,gambuzza2021stability}, here we harness topological signals on higher-order simplices, to garner insight into the interplay between topology and dynamics.

Despite these important scientific advances, higher-order networks have the limitation that they only consider monolayer simplicial complexes formed by a single higher-order network (layer), while a
 large variety of complex biological, social and technological systems are described by multilayer structures~\cite{bianconi2018multilayer,boccaletti2014structure,kivela2014multilayer}. 
 
 Multilayer networks~\cite{bianconi2018multilayer,boccaletti2014structure,kivela2014multilayer}
  are formed by  a set of nodes  linked to each other by different layers indicating interactions of different types.  The research on multiplex networks has been rapidly evolving in recent years and it has been shown that multiplex networks  display a rich interplay between structural and dynamical properties affecting diffusion~\cite{sole2013spectral,gomez2013diffusion,radicchi2013abrupt}, epidemic spreading~\cite{de2016physics,de2018fundamentals}, percolation~\cite{buldyrev2010catastrophic,berezin2015localized,danziger2019dynamic,cellai2016message,radicchi2017redundant} and synchronization~\cite{del2016synchronization,ghosh2016birth,nicosia2017collective}. However, so far, with a few exceptions \cite{ferraz2021phase,sun2023dynamic,sun2021higher,ghorbanchian2022hyper}, most of the prior works have considered multiplex networks with exclusively pairwise interactions.

In this work, we formulate a topological  theory of higher-order multiplex simplicial complexes, where each layer of the multiplex network includes higher-order interactions (interactions between two or more nodes) of different types. We define the multiplex Hodge Laplacians and Dirac operators. These operators capture higher-order diffusion on multiplex simplicial complexes whose definition exploits the ubiquitous presence of the  overlap of the simplices ~\cite{bianconi2013statistical,menichetti2014weighted,musmeci2017multiplex,cellai2016message,radicchi2017redundant,bentley2016multilayer,battiston2017multilayer}, i.e. the simultaneous presence of many-body interactions of different types between the same set of nodes. In particular, we define metric matrices that couple the layers, and use them to define the multiplex boundary operator, multiplex cochains, and multiplex Hodge Laplacians. We characterize the spectral properties of multiplex Hodge Laplacians and show that these operators obey Hodge decomposition, allowing the investigation of solenoidal, irrotational and harmonic signals on multiplex simplicial complexes. Additionally, we characterize the properties of the resulting higher-order diffusion on random  and real multiplex simplicial complex datasets.
Our results show that the considered topological coupling between the layers  can have opposing effects depending on the topology of the multiplex simplicial complex. Indeed, in some cases it can speed up the dynamics of higher-order topological signals, while in others, it can slow it down.
This has potential applications in control \cite{d2023controlling}, where one can modulate the speed of dynamic processes by changing the overlap of simplices, even if only interactions on one layer of the multiplex simplicial complex can be externally modified. Such approaches may be useful in studying brain disorders like schizophrenia where circular inference (or solenoidal signals representing circular beliefs) is though to underlie pathological behavior \cite{jardri2013circular,jardri2017experimental}.

The results provided on this work may be relevant to a variety of complex systems displaying higher-order interactions of different type, including brain networks which are known to be both multilayer and higher-order 
\cite{reimann2017cliques,petri2014homological,santoro2023higher,faskowitz2022edges,santos2019topological,bentley2016multilayer,presigny2022colloquium,battiston2017multilayer,nurisso2023unified}. 
It is worth noting that multiplex pairwise networks are a special case of multilayer simplicial complexes formed exclusively by 0-simplices (nodes) and 1-simplices (links). As such, the results illustrated in this work on multiplex simplicial complexes extend to multilayer pairwise networks as well, however the introduction of the multilayer Hodge Laplacian allows us to investigate the topology of the system by considering higher-order simplices.

The paper is organized as follows.
In Sec. 2 we introduce the key aspects of the topology of multiplex simplicial complexes; in Sec.3 we define the multiplex Hodge Laplacian and multiplex Dirac operators; in Sec.4 we characterize the higher-order diffusion dynamics of multiplex topological signals; in Sec. 5 we  discuss our numerical results on synthetic and real multiplex simplicial complexes. Finally in Sec. 7 we  provide the concluding remarks.
The paper is enriched by an extensive Appendix providing a mathematically rigorous account of the topology of multiplex simplicial complexes introduced and used in the main body of the paper.

\begin{figure}
\begin{center}
\includegraphics[width=\columnwidth]{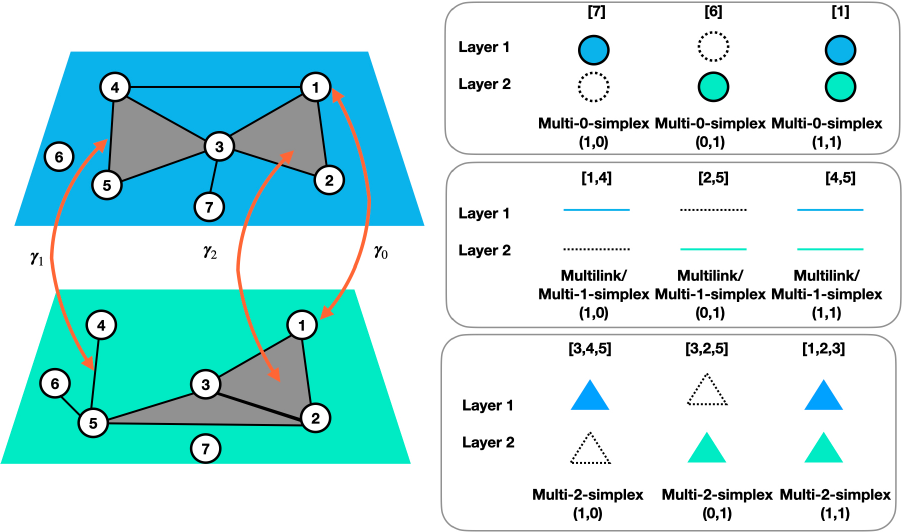}
\end{center}
\caption{{\bf Multiplex simplicial complexes and multisimplices.} A multiplex simplicial complex formed by $M=2$ layers and $\bar{N}=7$ nodes is shown (on the left) together with a schematic description of multi-simplices (on the right) including multi-0-simplices (multinodes), multi-1-simplices (multilinks) and multi-2-simplices (multitriangles).} 
\label{fig:1}
\end{figure}
\section{Topology of multiplex simplicial complexes}
\subsection{Multiplex simplicial complexes and multisimplices}
A multiplex network ~\cite{bianconi2018multilayer,boccaletti2014structure,kivela2014multilayer}  captures the pairwise interactions of different type present in a complex system and consists of a network of networks (layers) where  the nodes of each layer  are in one-to-one correspondence.

Here we define  multiplex simplicial complexes encoding  higher-order interactions of different types.
A multiplex simplicial complex $\mathcal{K}$ (see Figure $\ref{fig:1}$) is  the union of $M$ simplicial complexes $K^{[\alpha]}$ (with $\alpha\in \{1,2,\ldots,M\}$) each one determining a distinct layer of the multiplex simplicial complex. In neuroscience, this construction can, for instance, allow  to distinguish between different types of higher-order interactions between the same regions of the brain.

 Each simplicial complex $K^{[\alpha]}$ is formed by a set of simplices  indicating higher-order interactions of the same type existing among the same set of $\bar{N}$ nodes. An $n$-dimensional simplex $\sigma^{[\alpha]}\in K^{[\alpha]}$ is set of $n+1$ nodes describing a $n$-th order interaction of type $\alpha$.  For instance, a $0$-simplex is a node, a $1$-simplex is a link, a $2$-simplex is a triangle and so on.  A face of a $n$-dimensional simplex $\sigma^{[\alpha]}$ is a simplex of dimension  $n'$ (with $0\leq n'<n$) formed by a proper subset of the nodes of $\sigma^{[\alpha]}$. Every simplicial complex $K^{[\alpha]}$ is   closed under the inclusion of its faces, i.e. if a simplex $\sigma^{[\alpha]}$ belongs to $K^{[\alpha]}$, then all its faces  belong to the simplicial complex as well. We indicate with $N_n^{[\alpha]}$ the number of $n$-dimensional simplices present in layer $\alpha$ and with $N_n=\sum_{\alpha=1}^M N_n^{[\alpha]}$ the total number of $n$-dimensional  simplices in the multiplex  simplicial complex.
Moreover we  define  the dimension $d$ of the multiplex simplicial complex as the largest dimension of its simplices.

An important property of multiplex networks, that here we extend to multiplex simplicial complexes,  is that the nodes of different layers are in  one to one correspondence. In particular, corresponding nodes belonging to different layers are called replica nodes.  In the higher-order setting, this notion extends to higher-order simplices, and one can therefore speak of replica simplices, i.e. replica links, replica triangles and so on. 

 {In order to define replica simplices let us introduce the aggregated simplicial complex $\bar{K}$ that encodes all interactions of the associated multiplex simplicial complex, regardless of their type. Specifically, the aggregated simplicial complex is formed by the a set of  simplices $\bar{\sigma}$ where each simplex $\bar{\sigma}\in \bar{K}$ indicates  the presence of at least one type of  higher-order interaction between its constituents nodes.}
The replica simplices of the multiplex simplicial complex are representations of the simplices $\bar{\sigma}$ that  belong to the aggregated simplicial  complex $\bar{K}$.
Thus, each node $\bar{\sigma}=[v_0]\in \bar{K}$ is represented by the replica nodes $\bm\sigma=(\sigma^{[1]},\sigma^{[2]},\ldots, \sigma^{[M]})$, where $\sigma^{[\alpha]}$ indicates the replica node $\sigma^{[\alpha]}=[v_0;\alpha]$ which is the  representation of node $v_0$  in layer $\alpha$. Similarly, each higher dimensional simplex $\bar{\sigma}=[v_0,v_1\ldots,v_n]$ among $n$ nodes  can be  represented as a vector of replica simplices $\bm\sigma=(\sigma^{[1]},\sigma^{[2]},\ldots, \sigma^{[M]})$, where $\sigma^{[\alpha]}=[v_0;\alpha,v_1;\alpha\ldots v_n;\alpha]$ represents the simplex $\bar{\sigma}$ in layer $\alpha$. Note that replica simplices are `potential simplices' defined on all layers, but not all replica simplices may exist on all layers. In order to define the actual structure of a multiplex simplicial complex and determine which replica simplices are present in it, we have two options: either we specify the set of all the replica simplices $\sigma^{[\alpha]}\in K^{[\alpha]}$ present in  every layer $\alpha$, or we make use of the  the notion of multisimplices that we will introduce in the following. As we will see, this latter representation of the multiplex simplicial complex has the advantage that it explicitly keeps track of the overlap among the replica simplices.

Link overlap is a very important and well known property of real multiplex networks which refers to the case where two nodes are connected in more than one layer.
Link overlap is described by multilinks~\cite{bianconi2013statistical,menichetti2014weighted,musmeci2017multiplex,bentley2016multilayer,cellai2016message,radicchi2017redundant,battiston2017multilayer},  which capture all the possible ways in which two nodes can be connected in different layers. In order to capture this important property in our framework,  we extend  the notion of multilinks on multiplex networks to multisimplices on multiplex simplicial complexes.

While on a monoplex simplicial complex each simplex can be either be present or not, in a multiplex simplicial complex, any higher-order interaction (simplex) among $n$ nodes can independently be present or not in each  different layer.
In order to account for all the possible types of interactions  between a given set of $n$ nodes, we associate to each set of replica simplices $\bm\sigma=(\sigma^{[1]},\sigma^{[2]},\ldots, \sigma^{[M]})$ a vector $\vec{m}_{{\sigma}}=(m^{[1]}_{{\sigma}},m^{[2]}_{{\sigma}},\ldots, m^{[M]}_{{\sigma}})$ that captures whether the simplex  $\sigma^{[\alpha]}$ is present in  layer $\alpha$ ($m_{{\sigma}}^{[\alpha]}=1$) or not ($m^{[\alpha]}_{{\sigma}}=0$).
In this way we define multisimplices of type $\vec{m}$  representing every possible  given pattern of interactions among the same set of nodes across the different layers. In the case of two layers we can hence observe non trivial multi-0-simplices (multinodes) of type $\vec{m}$ given by $(1,0), (0,1)$ and $(1,1)$ indicating whether the node is present (i.e. it is connected) only in the first, only in the second or in both layers. Moreover we can observe multi-1-simplices (multilinks) and multi-2-simplicies (multitriangles) of type $\vec{m}$ given by $(1,0),(0,1)$ and $(1,1)$ indicating whether links or triangles are present only in the first, only in the second or in both layers respectively (see Figure $\ref{fig:1}$). Hence, the multisimplices give direct and complete information about the overlap among replica simplices across different layers.

In multilayer network literature~\cite{sole2013spectral,gomez2013diffusion,radicchi2013abrupt},  diffusion among the layers of the multiplex network is often treated with interlinks, i.e. links connecting the replica nodes in different layers. However interlinks do not have a clear and natural higher-order generalization.
Therefore here we do not use interlinks; rather, we harness multisimplices to define a way to topologically couple the layers of the multiplex simplicial complexes.  
To this end, we extend weighted algebraic topology operators \cite{baccini2022weighted,meng2020weighted,horak2013spectra}  to describe the topology of multiplex simplicial complexes. We present a brief account of our framework based on  homology and Hodge theory here.
For our full mathematical framework, we refer the reader to our Appendix.

\subsection{Multiplex cochains, and coboundary operators} 
Topological signals are dynamical variables sustained by the simplices of a simplicial complex. Typically topological signals sustained by $n$-dimensional simplices are represented as $n$-cochains.
In a multiplex network, we consider multiplex $n$-cochains $f\in C^n$, where $C^n$ is the set of all multiplex cochains, uniquely defined by the  $N_n$ dimensional vector ${\bf f}$ taking a real value on each $n$-dimensional replica simplex present in the multiplex simplicial complex.
In particular, choosing a basis in which the $n$-dimensional simplices are listed consecutively for different layers, we have 
\bea
{\bf f}=\left(\begin{array}{c}{\bf f}^{[1]}\\{\bf f}^{[2]}\\\vdots\\{\bf f}^{[M]}\end{array}\right),
\eea
where ${\bf f}^{[\alpha]}$ is a $N_n^{[\alpha]}$ dimensional vector defined on each of  the $n$-dimensional replica simplex $\sigma^{[\alpha]}$ present on layer $\alpha$, i.e. with $\sigma^{[\alpha]}\in K^{[\alpha]}$.

In order to define the topology of multiplex simplicial complexes, generalizing the topology of monolayer simplicial complexes~\cite{hatcher2005algebraic}, our first step is to introduce the multiplex boundary operator and coboundary operators.
Here, we adopt the simple approach of defining the multiplex  coboundary operator as the direct sum of the  coboundary operators of each individual layer (see Appendix).
Thus, we have that the multiplex coboundary operator $\delta_{n-1}$ can be expressed in matrix form by the $n$-th coboundary matrix 
 ${\mathcal {B}}_{n}$ of  block diagonal form given by 
\bea
{\mathcal {B}}_{n}=\left(\begin{aligned}& {B}^{[1]}_{n} && 0 && 0 && 0\\& 0 && {B}^{[2]}_{n} &&0 &&0 \\  & 0 && 0  && \ldots &&0 \\& 0 && 0  && 0 && {B}^{[M]}_{n} 
\end{aligned}\right),
\eea
where ${B}^{[\alpha]}_{n}$ is the 
$n$-coboundary matrix matrix for layer $\alpha$.
For instance, for a  2-layer multiplex simplicial complex,  the  $n$-th coboundary matrix is  given by 
\bea
{\mathcal {B}}_{n}=\left(\begin{aligned}& {B}^{[1]}_{[n]} && 0\\& 0 && {B}^{[2]}_{n}  \end{aligned}\right).
\eea
The multiplex boundary operator $\partial_n$ is instead captured by the boundary matrix $\mathcal{B}_n^{\top}$.
An important property of the boundary operators, which is also shared by our definition of multiplex operators, is that ``the boundary of the boundary is null", i.e.
\bea
{\mathcal {B}}_{n}^{\top}{\mathcal {B}}_{n+1}^{\top}&=&0,\nonumber \\
{\mathcal {B}}_{n+1}{\mathcal {B}}_{n}&=&0,
\label{bb}
\eea
for any $n\geq 1$ (see Appendix for details).
\subsection{Metric matrices and coupling among different layers}
\label{sec:metric}
So far, we have have described  multiplex simplicial complexes as  simplicial complexes built by the union of all the simplices present in every layer. However, in this way the simplices of different layers are uncoupled.
In order to couple the layers, we leverage on the ubiquitous presence of  the overlap of the simplices  and we introduce metric matrices that couple the overlapping simplices among the layers of the multiplex simplicial complexes.

The metric matrices ${\mathcal{G}}_n^{-1}$ define non-degenerate scalar products between $n$-dimensional multiplex cochains.
In particular, given $f_1,f_2\in C^{n}$, their scalar product is defined as 
\bea
\Avg{f_2,f_1}={\bf f}_2^{\top}\mathcal{G}_{n}^{-1}{\bf f}_1,
\eea
where $\mathcal{G}_n$ are $N_n\times N_n$  symmetric, and positive definite matrices defined among the $n$-dimensional replica simplices present in the simplicial complex $\mathcal{K}$.
Typically, for a monoplex simplicial complex, the metric matrices are taken to be diagonal, with diagonal elements given by the affinity weight of the corresponding $n$-dimensional simplex.
Here, in order to couple the different layers of the multiplex simplicial complex, we consider  non-diagonal metric matrices $\mathcal{G}_n$ that enforce a coupling between overlapping simplices  modulated by the parameter $\gamma_n$. 

 {Since the metric matrices will be defined through the multisimplices, let us, for ease of notation, introduce the auxiliary metric matrices $\tilde{\mathcal{G}}_n$ defined among all the possible $n$-dimensional replica simplices of the multiplex simplicial complex (i.e. replica simplices either present and non-present in $\mathcal{K}$). 
These auxiliary metric matrices $\tilde{\mathcal{G}}_n$ have rank $N_n$ and are  such that their action  reduces to the action of the metric matrices ${\mathcal{G}}_n$ on the  replica simplices present in the simplicial complex. 
The elements of the matrix $\tilde{ \mathcal G}_{n}$ can be non-zero only among replica simplices. Among any two $n$-dimensional  replica  simplices $\sigma^{[\alpha]}$ and $\sigma^{[\alpha^{\prime}]}$ belonging to the multisimplex $\vec{m}_{{\sigma}}=(m^{[1]}_{{\sigma}},m^{[2]}_{{\sigma}},\ldots, m^{[M]}_{{\sigma}})$, the auxiliary metric matrices have elements: 
\bea
\tilde{\mathcal G}_{n}(\sigma^{[\alpha]},\sigma^{[\alpha^{\prime}]})=\left\{\begin{array}{lll}({W_n}/{\mathcal{C}})m_{\sigma}^{[\alpha]}& \mbox{if}& \alpha=\alpha^{\prime},\\
({W_n}/{\mathcal{C}})\gamma_n m_{\sigma}^{[\alpha]}m_{\sigma}^{[\alpha^{\prime}]} & \mbox{if}& \alpha\neq \alpha^{\prime},\end{array}\right.
\label{def:Gn}
\eea
 with the parameters $\gamma_n$ and $W_n$ taking values in the range $\gamma_n\in [0,1)$ and  $W_n>0$. 
 Given the auxiliary metric matrix $\tilde{\mathcal{G}}_n$, the metric matrix $\mathcal{G}_n$ can be obtained as the  submatrix of $\tilde{\mathcal{G}}_n$ acting on linear span of the set of all replica simplices present on  the multiplex simplicial complex. [Note that the submatrix of $\tilde{\mathcal{G}}_n$ acting on all the replica simplices not present in the simplicial complex turns out to be null.]
From Eq. (\ref{def:Gn}) it is clear that the parameters $\gamma_n$ determine the coupling between the  layers for $\gamma_n\neq 0$ while   when $\gamma_n=0$ for all values of $n$,  the layers of the multiplex simplicial complex are completely uncoupled.
 Moreover, $W_n>0$ are real parameters that can be used to tune the scale of the metric matrices for different dimensions $n$.
Moreover, in  Eq. (\ref{def:Gn}), $\mathcal{C}$ indicates the normalization constant that ensures that  the matrix $\mathcal{G}_{n}$ has determinant equal to $W_n^{N_n}>0$.
This constant  is expressed in terms of $\mu=\sum_{\alpha=1}^M m^{[\alpha]}$ indicating  the multiplicity of overlap of the $n$-dimensional simplex $\bar{\sigma}$, i.e. the number of layers in which the simplex $\bar{\sigma}$ is present.
It follows that  the constant $\mathcal{C}$ is given by
\bea
\mathcal{C}=[(1-\gamma_n+\mu\gamma_n)(1-\gamma_n)^{\mu-1}]^{1/\mu}.
\eea}
 {It is instructive to express the matrix elements of $\tilde{\mathcal{G}}_n$ for the simple case of overlapping replica simplices $\bm\sigma=(\sigma^{[1]},\sigma^{[2]})$ in a multiplex simplicial complex with $M=2$  layers,  
\bea
\tilde{\mathcal{G}}_{n}(\bm{\sigma},\bm\sigma)=\frac{W_n}{\sqrt{1-\gamma_n^2m_{\sigma}^{[1]}m_{\sigma}^{[2]}}}\left(\begin{array}{cc}m_\sigma^{[1]}& \gamma_{n}m_{\sigma}^{[1]}m_{\sigma}^{[2]} \\ \gamma_n m_{\sigma}^{[1]}m_{\sigma}^{[2]} & m_{\sigma}^{[2]} \end{array}\right). \label{def:gk2}
\eea
}
 {From this expression is evident that if the two simplices overlap, (i.e. $\vec{m}_{\sigma}=(1,1)$) this matrix has rank $2$, a determinant $W_n^2$ and  two positive eigenvalues $W_n\sqrt{(1+\gamma_n)/(1-\gamma_n)}>0$ and $W_n\sqrt{(1-\gamma_n)/(1+\gamma_n)}>0$. Otherwise, if one of the two simplices is not present, (i.e. $\vec{m}_{\sigma}=(1,0)$ or $\vec{m}_{\sigma}=(0,1)$), this matrix  has rank $1$ with the non-zero eigenvalue $W_n>0$ corresponding to an eigenvector aligned to the simplex  present in the multiplex simplicial complex. Given that only overlapping replica simplices are coupled together,  it follows that the metric matrices $\mathcal{G}_n$,  acting exclusively on the space of replica simplices  present in the simplicial complex, are positive definite.}

For three layers, i.e. $M=3$, indicating with   $\bm\sigma$  the set of replica simplices $\bm\sigma=(\sigma^{[1]},\sigma^{[2]},\sigma^{[3]})$ we have instead,
\bea
\tilde{\mathcal{G}}_{n}(\bm\sigma,\bm\sigma)=\frac{W_n}{[(1-\gamma_n+\mu\gamma_n)(1-\gamma_n)^{\mu-1}]^{1/\mu}}\left(\begin{array}{ccc}m_{\sigma}^{[1]}& \gamma_{n}m_{\sigma}^{[1]}m_{\sigma}^{[2]} &\gamma_{n}m_{\sigma}^{[1]}m_{\sigma}^{[3]}\\ \gamma_n m_{\sigma}^{[1]}m_{\sigma}^{[2]} & m_{\sigma}^{[2]} &  \gamma_{n}m_{\sigma}^{[2]}m_{\sigma}^{[3]}\\ \gamma_{n}m_{\sigma}^{[3]}m_{\sigma}^{[1]}&  \gamma_{n}m_{\sigma}^{[2]}m_{\sigma}^{[3]} & m_{\sigma}^{[3]}\end{array}\right).
\eea
where $\mu=\sum_{\alpha=1}^M m_{\sigma}^{[\alpha]}$ is the multiplicity of overlap of the simplex.
 {Following similar arguments as in the case $M=2$,  it can be easily seen  that also for $M=3$, the matrice $\mathcal{G}_n$, acting on the space of replica simplices present in the multiplex simplicial complex, are positive definite.}

\section{Multiplex Hodge Laplacians and multiplex Dirac operator}

Having defined the metric matrices among multiplex cochains, one can now define the adjoint of the coboundary operator that will be instrumental in defining the multiplex Hodge Laplacians and Dirac operators. 
In matrix form, the multiplex adjoint coboundary operator $\delta_{n-1}^{\star}$ is expressed (see Appendix for details) by the matrix $\mathcal{B}_{n}^{\star}$ which depends on the multiplex coboundary matrix $\mathcal{B}_{n}$ and by the metric matrices $\mathcal{G}_{n}$ and $\mathcal{G}_{n-1}$ as
\bea
\mathcal{B}_{n}^{\star}=\mathcal{G}_{n-1}\mathcal{B}_{n}^{\top}\mathcal{G}_{n}^{-1}.
\label{Bstar}
\eea

We are now in a position to define the  multiplex $n$-Hodge Laplacians that generalize the $n$-Hodge Laplacian of monoplex simplicial complexes~\cite{eckmann1944harmonische,horak2013spectra,jost2019hypergraph,lim2020hodge,meng2020weighted,mulas2022graphs}, and  characterize the diffusion of higher-order topological signals from $n$-dimensional multisimplices to $n$-dimensional multisimplices through ($n+1$)-dimensional and ($n-1$)-dimensional multisimplices.   In matrix form the multiplex Hodge Laplacian is expressed as
\bea
\mathcal{L}_{n}=\mathcal{L}_{n}^{up}+\mathcal{L}_{n}^{down}
\eea
with 
\bea
\mathcal{L}_{n}^{up}&=&\mathcal{B}_{n+1}^{\star}\mathcal{B}_{n+1}=\mathcal{G}_n\mathcal{B}_{n+1}^{\top}\mathcal{G}_{n+1}^{-1}\mathcal{B}_{n+1},
\nonumber \\
\mathcal{L}_{n}^{down}&=&\mathcal{B}_{n}\mathcal{B}_{n}^{\star}=\mathcal{B}_{n}\mathcal{G}_{n-1}\mathcal{B}_{n}^{\top}\mathcal{G}_{n}^{-1},
\label{Laplacian_up_down}
\eea
where $\mathcal{L}_0^{down}=0$.
Therefore, in a $d=2$ dimensional multiplex simplicial complex (in which each layer contains nodes, links, triangles, we have the following Hodge Laplacian operators:
\bea
\mathcal{L}_{0}=\mathcal{G}_0\mathcal{B}_{1}^{\top}\mathcal{G}_1^{-1}\mathcal{B}_{1},\nonumber \\
\mathcal{L}_{1}^{down}=\mathcal{B}_{1}\mathcal{G}_0\mathcal{B}_{1}^{\top}\mathcal{G}_1^{-1},\nonumber \\
\mathcal{L}_{1}^{up}=\mathcal{G}_1\mathcal{B}_{2}^{\top}\mathcal{G}_2^{-1}\mathcal{B}_{2}, \nonumber \\
\mathcal{L}_{2}^{down}=\mathcal{B}_{2}\mathcal{G}_1\mathcal{B}_{2}^{\top}\mathcal{G}_2^{-1}.
\label{eq:multiplexhodge2}
\eea
Interestingly, it can be easily proven that these multiplex Hodge-Laplacians, although asymmetric, have a real, non-negative spectrum (see Appendix).\\ 
Moreover, from the definition of the multiplex Hodge Laplacians we notice  that since the boundary and the coboundary matrices obey Eq.$(\ref{bb})$,   the multiplex Hodge Laplacians satisfy 
\bea
\mathcal{L}_{n}^{up}\mathcal{L}_{n}^{down}=\mathcal{L}_{n}^{down}\mathcal{L}_{n}^{up}=0,
\eea
which implies Hodge decomposition.
 A consequence of Hodge decomposition  is  that not only do $\mathcal{L}_n^{up}$ and $\mathcal{L}_n^{down}$ commute, but also for any left eigenvector of  the multiplex Hodge Laplacian of $\mathcal{L}_{n}$, corresponding to the eigenvalue $\lambda$, there are only three options:
\begin{itemize}
\item[(1)] If $\lambda=0$ the eigenvector is in both a left eigenvector in the kernel of $\mathcal{L}_n^{up}$ and a left eigenvector in the kernel of $\mathcal{L}_n^{down}$. 
\item[(2)] If $\lambda=0$ only two mutually excluding options exists: 
\begin{itemize}\item[(i)]the left eigenvector is also a left eigenvector of  $\mathcal{L}_n^{up}$  corresponding to eigenvalue $\lambda$ and simultaneously is a left eigenvector of $\mathcal{L}_n^{down}$ corresponding to an eigenvalue zero,
\item[(ii)] the eigenvector is also a left eigenvector of  $\mathcal{L}_n^{down}$  corresponding to eigenvalue $\lambda$ and simultaneously is a left eigenvector of $\mathcal{L}_n^{up}$ corresponding to the  zero eigenvalue,
\end{itemize}
\end{itemize}
Similar options apply for the right eigenvectors of $\mathcal{L}_n$.

An important spectral property of the multiplex $n$-Hodge Laplacian is that the dimension of its kernel is given by the multiplex $n$-Betti number, which is given by the sum of the  sum of the $n$-Betti numbers of each individual layer (see Appendix for details).

We also define the multiplex Dirac operators that generalize the Dirac operators of  simplicial complexes \cite{bianconi2021topological,baccini2022weighted,wee2023persistent}. Dirac operators are algebraic topological operators that are emerging as an important tool to couple topological signals of different dimensions~\cite{bianconi2023dirac,calmon2022dirac,giambagli2022diffusion,calmon2023local,calmon2023dirac}.
Extending the definition of the weighted Dirac operator \cite{baccini2022weighted}, here we define the multiplex Dirac operator.
The multiplex Dirac operator can be expressed as sums of partial Dirac operators acting exclusively on $(n-1)$ dimensional and $n$-dimensional signals,  $D_n:C^{n-1}\oplus C^n\to C^{n-1}\oplus C^n$ which in matrix form read
\bea
\mathcal{D}_n=\left(\begin{array}{cc}0& \mathcal{G}_{n-1}\mathcal{B}_{n}^{\top}\mathcal{G}_n^{-1}\\
	{\mathcal{B}_{n}}& 0\end{array}\right).
\eea
These partial Dirac operators can be used to define the  multiplex Dirac operator of the entire multiplex simplicial complex, coupling simplices of different dimensions. The Dirac operator acts on the direct sum of topological signals of any given  dimension, allowing cross-talk between different dimensions.
On a multiplex simplicial complex of dimension $d=2$  the  Dirac operator $D:C^0 \oplus C^1\oplus C^2\to C^0 \oplus C^1 \oplus C^2$  reads in matrix form as
\bea
\mathcal{D}=\left(\begin{array}{ccc}0& \mathcal{G}_{0}\mathcal{B}_{1}^{\top}\mathcal{G}_1^{-1} &0\\
	{\mathcal{B}_{1}}& 0 &\mathcal{G}_{1}\mathcal{B}_{2}^{\top}\mathcal{G}_2^{-1}\\
 0& {\mathcal{B}_{2}} &0
 \end{array}\right).
\eea
A key property of the  Dirac operator is that it can be interpreted as the ``square root" of the Hodge Laplacians. Indeed, on a $2$-dimensional multiplex simplicial complex we have
\bea
\mathcal{D}^2=\left(\begin{array}{ccc}\mathcal{L}_0 &0 &0\\
0&	{\mathcal{L}_{1}}& 0 \\
 0& 0& {\mathcal{L}_{2}}
 \end{array}\right).
\eea
Consequently, we obtain that the eigenvalues $\eta$ of the  Dirac operator  are given by the square root of the eigenvalues of the Hodge Laplacians taken with both positive and negative sign, i.e.
\bea
\eta=\pm \sqrt{\lambda},
\eea 
where $\lambda$ is a generic eigenvalue of any of the Hodge Laplacians of the simplicial complex. Thus, it follows that while the Hodge Laplacians are positive semi-definite, the Dirac operator is not.
\begin{figure}
\begin{center}
\includegraphics[width=\columnwidth]{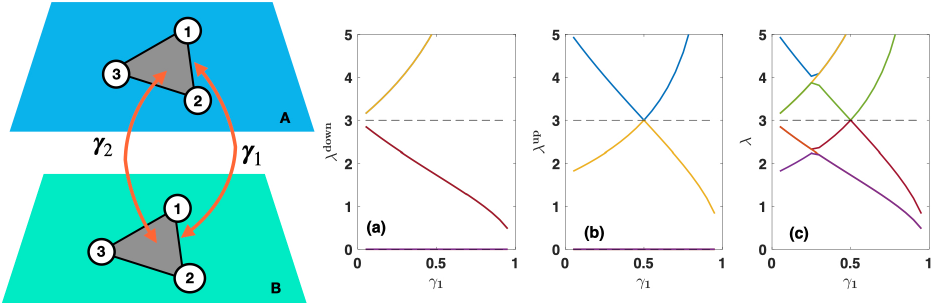}
\end{center}
\caption{{\bf Example of $d=2$-dimensional multiplex simplicial complexes and its spectrum.} A simple multiplex simplicial complex is shown (on the left). On the right, the spectrum of the $\mathcal{L}_1^{down},\mathcal{L}_1^{up}$ and $\mathcal{L}_1$ Hodge Laplacians. All the  $\lambda^{down}$  of $\mathcal{L}_1^{down}$ in (panel (a)), $\lambda^{up}$ of $\mathcal{L}_1^{up}$  (panel (b)), and $\lambda$ of $\mathcal{L}_1$ (panel (c)) respectively are displayed as a function of $\gamma_1$ for $\gamma_2=0.5$. The eigenvalue curves are color coded by eigenvalue, and overlapping curves denote degenerate eigenvalues.} 
\label{fig:2}
\end{figure}

\begin{figure}
\begin{center}
\includegraphics[width=0.8\columnwidth]{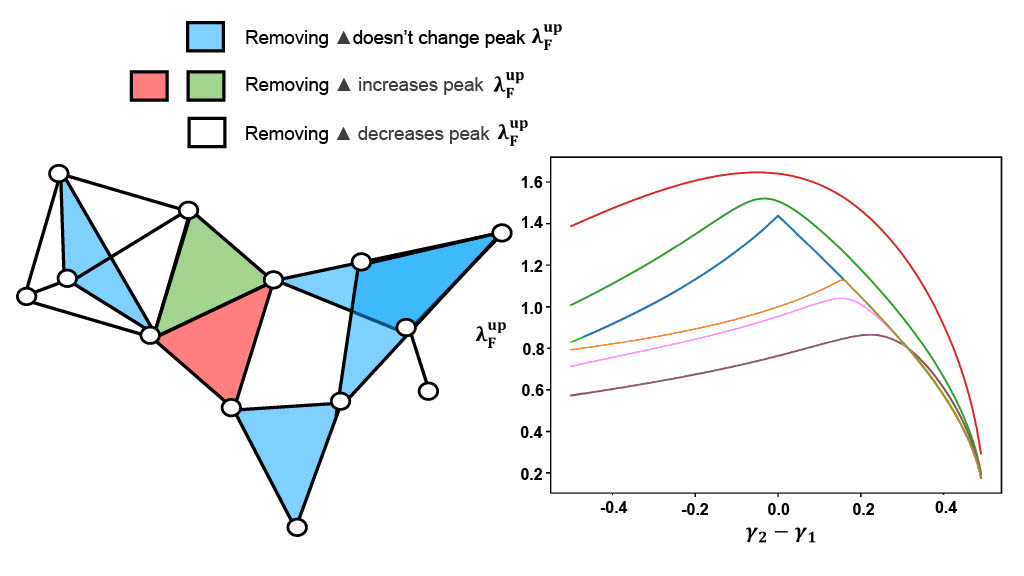}
\end{center}
\caption{{\bf Effect of overlap of triangles on the Fiedler eigenvalue of the $1$-up-Hodge Laplacian} An example of a  multiplex simplicial complex formed by $M=2$  layers with the same network skeleton (left). In the first layer, all triangles are filled, in the second layer we remove a single triangle at a time and compute the spectrum of $\mathcal{L}_{1}^{up}$. The corresponding Fiedler eigenvalues $\lambda_F^{up}$ are plotted  as a function of $\gamma_2- \gamma_1$ (right). Specifically, the removal of any of the blue triangles (removed one at a time) does not change the Fiedler eigenvalue curve. The removal of the red and the green triangle (one at a time) corresponds to an increase in Fiedler eigenvalue (corresponding the red and green curves on the right). Lastly, the removal of any of the six transparent triangles (one at a time) on the left, correspond to the three curves with the lowest peak Fiedler eigenvalue on the right (color coded in different colors). The symmetry in removal of the white triangles results in degeneracies in the Fiedler eigenvalues, resulting in three unique curves. The parameter $\gamma_1$ is kept fixed at $\gamma_1=0.5$ while $\gamma_2$ is varied.} 
\label{fig:3}
\end{figure}

\begin{figure}
\begin{center}
\includegraphics[width=0.9\columnwidth]{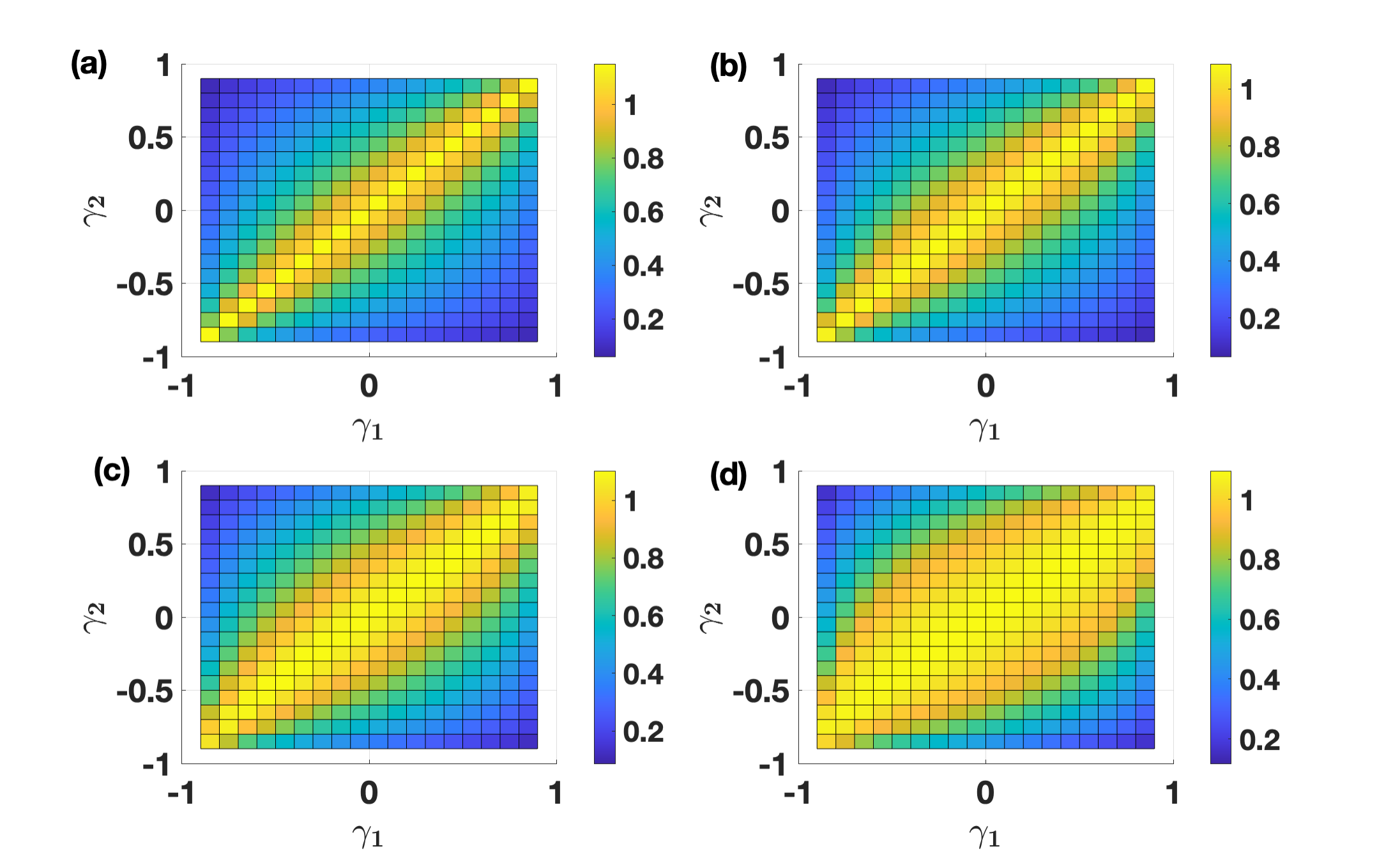}
\end{center}
\caption{{\bf The  Fielder eigenvalue ${\lambda}_F^{up}$ for a  random model with tunable triangle overlap}  {The heatmap of the Fiedler eigenvalue $\lambda_F^{up}$ of a random $2$-dimensional simplicial complex with tunable level of triangle overlap  is plotted as a function of $\gamma_1$ and $\gamma_2$. The random simplicial complex model is constructed  by drawing empty triangles with probability $p_2=12/\bar{N}^2$ between $\bar{N}=40$ nodes. The resulting clique complex forms the simplicial complex in the first layer. The second layer is constructed by considering each triangle present in the first layer and including it (together will all its faces) in the  the second layer  with probability $p_1$. The panels correspond to the results obtained  for $p_1=1$ (panel (a)), $p_1=0.75$ (panel(b)),  $p_1=0.5$ (panel (c)) and $p_1=0.25$ (panel (d)). The data is averaged over $10$ iterations of the model. We observe that for large overlap of the simplices, i.e. high values of $p_1$, the maximum of $\lambda_F^{up}$ for $\gamma_1\simeq \gamma_2$ is a robust feature of this model.} }
\label{fig:3b}
\end{figure}
\begin{table*}[htb!]
\begin{center}
\begin{tabular}{ccc}
MSC  & Layer 1&Layer 2\\
\hline
\hline
Microbiome A&Anterior nares&L Retroauricular crease\\
Microbiome B&Throat&Palatine Tonsils\\
Microbiome C&Throat&Tongue dorsum\\
Microbiome D&Hard palate&Tongue dorsum\\
C.elegans A&ElectrJ&MonoSyn\\
C.elegans B&ElectrJ&PolySyn\\
C.elegans C&MonoSyn&PolySyn\\
\hline\\
\end{tabular}\\
\begin{tabular}{cccc}
MSC & Layer 1&Layer 2 &Layer 3\\
\hline
\hline
Microbiome E&Anterior Throat&Palatine Tonsils&Tongue dorsum\\
C.elegans D&ElectrJ&MonoSyn&PolySyn\\
\hline
\end{tabular}
\end{center}
\caption{{\bf Layer composition of the considered multiplex  simplicial complexes (MSC).} We consider 9 multiplex simplcial complexes $7$  formed by $M=2$ layers (Layer 1 and Layer 2) and $2$ formed by $M=3$ layers (Layer 1, Layer 2 and Layer 3) of the microbiome multiplex network and the C.elegans connectome  (data  from \cite{manlio_github}).  The summary statistics of their  multisimplices are listed in Table $\ref{table1a}$ and Table $\ref{table2}$.}
\label{table1a}
\end{table*}

\begin{table*}[htb!]\begin{center}
\begin{tabular}{cccccccccc}
MSC &  $N^{(1,0)}$& $N^{(0,1)}$&$N^{(1,1)}$&$L^{(1,0)}$&$L^{(0,1)}$&$L^{(1,1)}$&$T^{(1,0)}$&$T^{(0,1)}$&$T^{(1,1)}$\\
\hline
\hline
Microbiome A&11&11&27&22&30&33&14&37&21\\
Microbiome B&12&14&70&106&147&336&429&490&968\\
Microbiome C&26&2&56&190&72&252&774&285&625\\
Microbiome D&11&12&46&61&144&180&127&550&358\\
C.elegans A&15&22&238&403&777&111&166&452&4\\
C.elegans B&1&26&252&352&1541&162&143&2112&27\\
C.elegans C&1&19&259&258&1073&630&218&1901&238\\
\hline
\end{tabular}
\end{center}
\caption{{\bf Major properties of the considered multiplex simplicial complexes (MSC) with $M=2$ layers.} We consider several duplex simplicial complexes formed by $M=2$ layers (Layer 1 and Layer 2 indicated in Table $\ref{table1a}$) of the microbiome multiplex network and the C.elegans connectome  (data  from \cite{manlio_github}). Each duplex simplicial complex includes  $N^{\vec{m}}$  multinodes of type $\vec{m}$, $L^{\vec{m}}$ multilinks of type $\vec{m}$, and $T^{\vec{m}}$ multitriangles of type $\vec{m}$. }
\label{table1b}
\end{table*}

\begin{table*}[htb!]
\begin{center}
\begin{tabular}{cccccccc}
\hline
\hline\\
MSC &$N^{(1,0,0)}$&$N^{(0,0,1)}$&$N^{(0,0,1)}$&$N^{(1,1,0)}$&$N^{(0,1,1)}$&$N^{(1,0,1)}$&$N^{(1,1,1)}$\\
\hline\\
Microbiome E&11&14&2&15&0&1&55\\
C.elegans D&0&0&4&1&22&15&237\\
\hline
\hline
\\
MSC &$L^{(1,0,0)}$&$L^{(0,0,1)}$&$L^{(0,0,1)}$&$L^{(1,1,0)}$&$L^{(0,1,1)}$&$L^{(1,0,1)}$&$L^{(1,1,1)}$\\
\hline\\
Microbiome E&83&128&53&107&19&23&229\\
C.elegans D&326&232&996&26&545&77&85\\
\hline
\hline\\
MSC&$T^{(1,0,0)}$&$T^{(0,0,1)}$&$T^{(0,0,1)}$&$T^{(1,1,0)}$&$T^{(0,1,1)}$&$T^{(1,0,1)}$&$T^{(1,1,1)}$\\
\hline\\
Microbiome E&340&427&222&434&63&89&534\\
C.elegans D&142&217&1877&1&235&24&31\\
\hline
\hline
\end{tabular}
\end{center}
\caption{{\bf Major properties of the considered multiplex simplicial complexes (MSC) of $M=3$ layers.} We consider two multiplex  simplicical complexes  formed by $M=3$ layers: the first one is built from the the microbiome multiplex network and the second one is built from the  C.elegans connectome  (data  from \cite{manlio_github}). Each multiplex  simplicial complex  includes  $N^{\vec{m}}$ multinodes of type $\vec{m}$, $L^{\vec{m}}$ multilinks of type $\vec{m}$, and $T^{\vec{m}}$ multitriangles of type $\vec{m}$. 
The three layers of the multiplex simplicial complexes are listed in Table $\ref{table1a}$.}
\label{table2}
\end{table*}

\section{Higher-order multiplex diffusion dynamics}

Having defined the multiplex Hodge Laplacian we can now investigate multiplex higher-order diffusion, extending the notion of higher-order diffusion in monoplex simplicial complexes~\cite{torres2020simplicial,ziegler2022balanced,muhammad2006control,schaub2020random,krishnagopal2021spectral} 
We consider the topological signal  ${\bf X}\in C^n$ encoded by a multiplex cochain. 
For the multiplex simplicial complex, multiplex higher-order diffusion of a $n$ topological signal $\bf X$ is given by 
\bea
\frac{d{\bf X}}{dt}=-{\mathcal L}_{n}{\bf X}.
\eea
Since the multiplex Hodge Laplacian obeys Hodge decomposition,  every $n$ topological signal ${\bf X}$ can be decomposed in a unique way into 
\bea
{\bf X}={\bf X}_1+{\bf X}_{2}+{\bf X}^{\text{harm}},
\eea
where we call ${\bf X}_1$ the irrotational component of the signal, ${\bf X}_2$ its solenoidal component, and ${\bf X}^{\text{harm}}$ its harmonic component, i.e.,
\bea
{\bf X}_1\in \mbox{im}({\bf L}^{down})\quad {\bf X}_2\in \mbox{im}({\bf L}^{up}),\quad{\bf X}^{\text{harm}}\in \mbox{ker}({\bf L}).
\eea
{This is analogous to the case of the monoplex $1-$Hodge Laplacian, where Hodge decomposition results in three orthogonal signals that are indeed curl-free (irrotational), divergence-free (solenoidal), and both curl-free and divergence-free (harmonic).}
Note, however, that due to the presence of the non-zero metric matrices, the irrotational and the soleinodal components of the multiplex network signals may not in general be irrotational and solenoidal on each individual layer. 

From the Hodge decomposition of the multiplex topological signals, it follows that  the multiplex higher-order diffusion can be captured by a set of three uncoupled equations: 
\bea
\frac{d{\bf X}_1}{dt}&=&-{\mathcal L}_{n}^{down}{\bf X}_1,\nonumber \\
\frac{d{\bf X}_2}{dt}&=&-{\mathcal L}_{n}^{up}{\bf X}_2,\nonumber\\
\frac{d{\bf X}^{\text{harm}}}{dt}&=&0.
\eea
Thus, the harmonic component is not affected by the diffusion dynamics and if present will remain unchanged in time. However, the irrotational component ${\bf X}_2$ and the solenoidal component ${\bf X}_2$ will independently relax to zero asymptotically in time with different time-scales. The timescale of diffusion of this form is controled by the leading eigenvalue of the Laplacian.
In particular we can 
 distinguish between two different Fiedler eigenvalues of the topological dynamics, one that describes the relaxation of the irrotational component $\lambda_F^{-}$ (and is the smallest non zero eigenvalue of  $\mathcal{L}_{n}^{down}$) and one that describes the relaxation to equilibrium of the solenoidal component $\lambda_F^{+}$ (and is the smallest non zero eigenvalue of  $\mathcal{L}_{n}^{up}$).
We have that  $\lambda_F^{-}$ is  a function of $\gamma_{n-1}$ and $\gamma_n$, and  $\lambda_F^{+}$ is  a function of $\gamma_n$ and $\gamma_{n+1}$. Thus, the two different eigenvalues can be independently controled by varying $\gamma_{n-1}$ and $\gamma_{n+1}$ respectively.
For each value of $n$, by monitoring  the dependence of the values of these two Fiedler eigenvalues  on the coupling constants, we can assess whether the introduced coupling between the layers speeds up or slows down the relaxation dynamics with respect to the case in which the layers are uncoupled ($\gamma_n=0$).
Moreover $\lambda_F^{-}$ is linear in $W_n$ while $\lambda_F^{+}$ if linear in $W_{n+1}$. Therefore their relative value can be tuned freely by changing the relative values of $W_n$ and $W_{n+1}$. This implies that by changing the values of the parameters $W_n$ and $W_{n+1}$ we can allow either the irrotational or the solenoidal component of the signal to relax faster.

We note that the Dirac operator can also be used to describe dynamics of topological signals. However, given that the Dirac operator is not positive definite, the appropriate dynamics should consider complex-values topological signals obeying the equation 
\bea
\frac{d{\bf Y}}{dt}=\textrm{i}{\mathcal D}{\bf Y},
\eea
where on a $d$-dimensional multiplex simplicial complex, ${\bf Y}\in \bigoplus_{n=1}^d C^n$ is a vector given by the direct sum of topological signals defined on replica simplices of every dimension.
This is a very interesting dynamics coupling signals on different dimensions, which may be relevant in the framework of continuous-time quantum walks and the recent interest in complex weights \cite{mulken2011continuous,bottcher2022complex,tian2023structural}. The above dynamics, clearly, will not relax to the harmonic eigenstates due to the presence of the imaginary unit on the right hand side. Therefore, the dynamics in this case will depend on the whole spectrum of the multiplex simplicial complex, rather than just on its Fiedler eigenvalues. The analysis of this dynamics requires detailed investigation, and may be a topic of future research.
 \begin{figure}
\begin{center}
\includegraphics[width=\columnwidth]{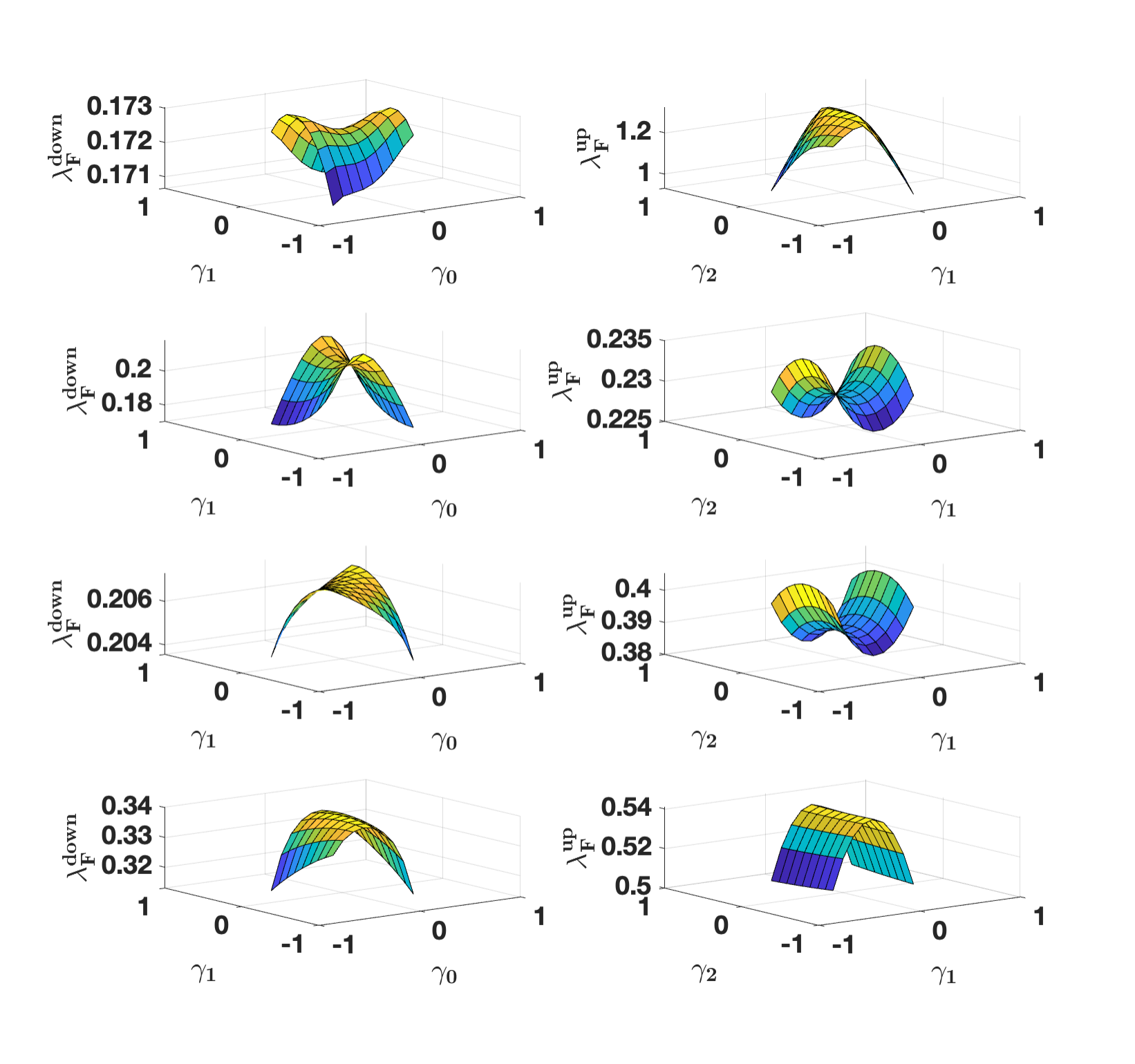}
\end{center}
\caption{{\bf Irrotational and solenoidal Fiedler eigenvalues as a function of the layer couplings for the Microbiome duplex simplicial complexes.} The irrotational and solenoidal Fiedler eigenvalues of $n=1$- multiplex Hodge Laplacians are plotted as a functions of the parameters $\gamma_n$ showing non trivial dependence with the coupling between the layers. The results are here shown from the top to the bottom for  the duplex simplicial complexes: Microbiome A, B, C, D whose layer composition and multisimplices statistics are reported in Table $\ref{table1a}$ and Table $\ref{table1b}$ respectively.} 
\label{fig:4}
\end{figure}
    \begin{figure}
\begin{center}
\includegraphics[width=\columnwidth]{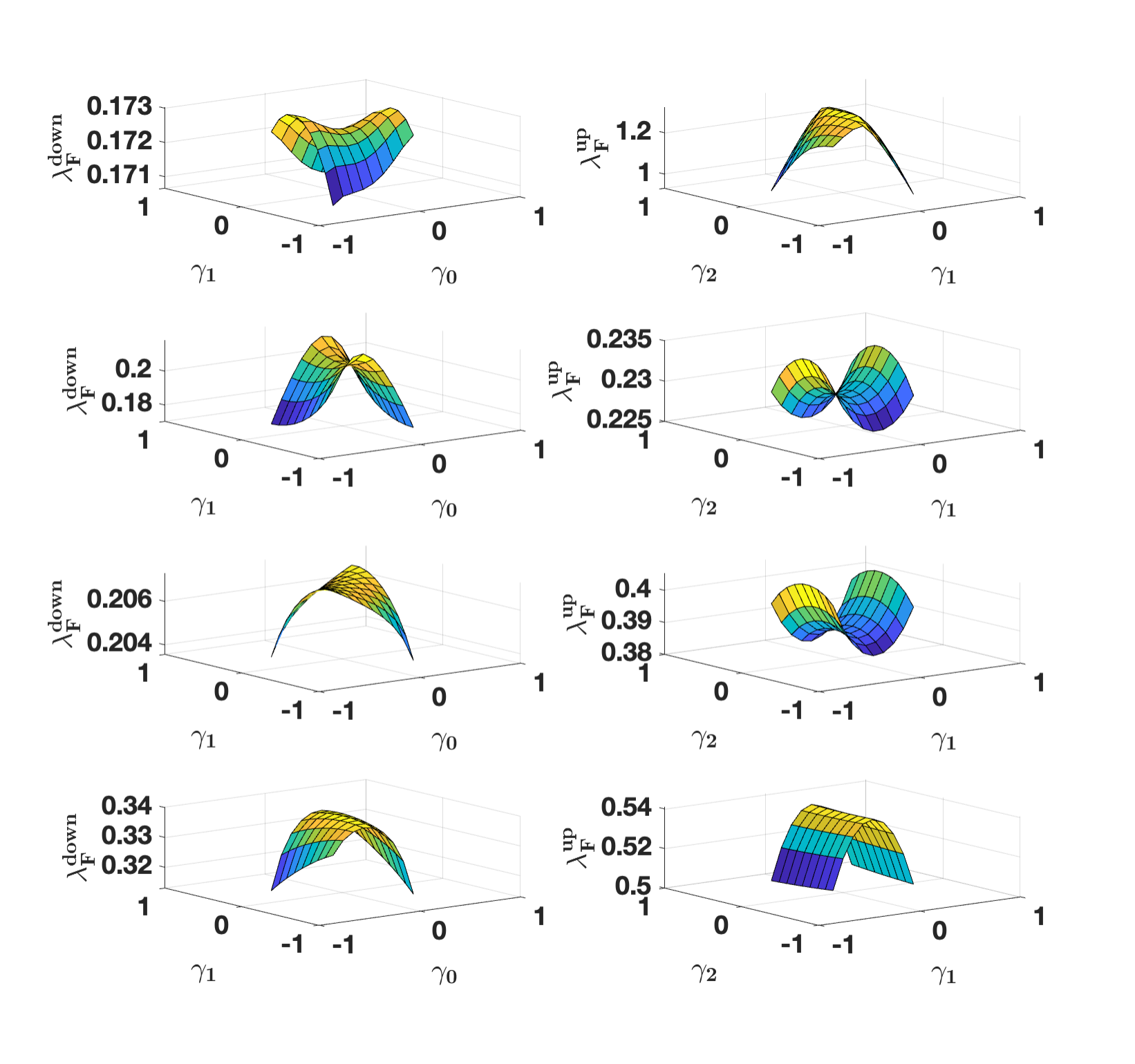}
\end{center}
\caption{{\bf Irrotational and solenoidal Fiedler eigenvalues as a function of the layer couplings for the C.elegans duplex simplicial complexes.} The irrotational and solenoidal Fiedler eigenvalues of $n=1$- multiplex Hodge Laplacians are plotted as a functions of the parameters $\gamma_n$ showing non trivial dependence with the coupling between the layers. The results are here shown from the top to the bottom for  the duplex simplicial complexes: C.elegans A, B, C whose layer composition and multisimplices statistics are reported in Table $\ref{table1a}$ and Table $\ref{table1b}$ respectively.} 
\label{fig:5}
\end{figure}

 \begin{figure}
\begin{center}
\includegraphics[width=\columnwidth]{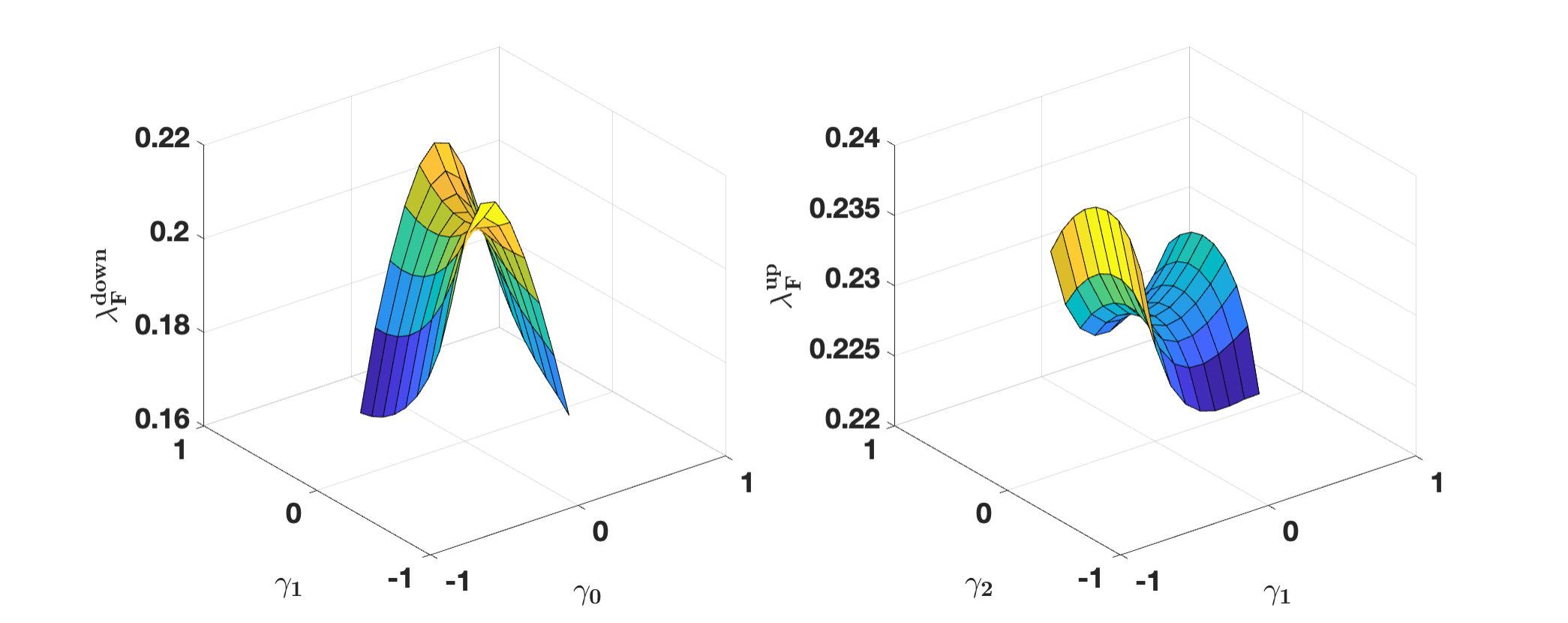}
\end{center}
\caption{{\bf Irrotational and solenoidal Fiedler eigenvalues as a function of the layer couplings for the Microbiome E multiplex simplicial complex.} The irrotational and solenoidal Fiedler eigenvalues of $n=1$- multiplex Hodge Laplacians are plotted as a functions of the parameters $\gamma_n$ showing non trivial dependence with the coupling between the layers. The results are here shown  for  the multiplex  simplicial complexes Microbiome E  whose layer composition and multisimplices statistics are reported in Table $\ref{table1a}$ and Table $\ref{table2}$ respectively.} 
\label{fig:6}
\end{figure}
 \begin{figure}
\begin{center}
\includegraphics[width=\columnwidth]{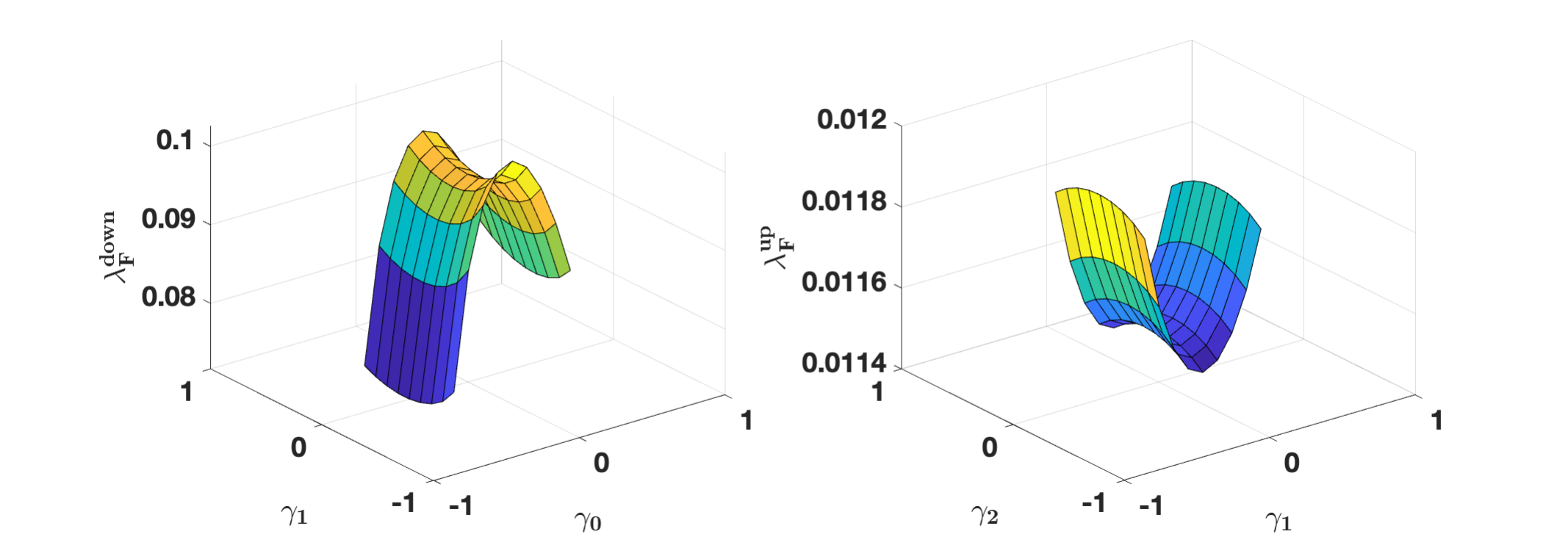}
\end{center}
\caption{{\bf Irrotational and solenoidal Fiedler eigenvalues as a function of the layer couplings for the C.elegans D multiplex simplicial complex.} The irrotational and solenoidal Fiedler eigenvalues of $n=1$- multiplex Hodge Laplacians are plotted as a functions of the parameters $\gamma_n$ showing non trivial dependence with the coupling between the layers. The results are here shown  for  the multiplex  simplicial complexes C.elegans D whose layer composition and multisimplices statistics are reported in Table $\ref{table1a}$ and Table $\ref{table2}$ respectively.} 
\label{fig:7}
\end{figure}

\section{Results on synthetic models and real data}
\subsection{Results on synthetic models}
We investigate the spectral and diffusion properties of some simple synthetic models of multiplex simplicial complexes.
As a first example we consider a multiplex simplicial complex of $M=2$ two layers and  $\bar{N}=3$ nodes,  formed by a filled triangle in each layer (see Figure $\ref{fig:2}$). The panels on the right plot all the eigenvalues (in different colors) of the $L_1^{down}$, $L_1^{up}$, and $L_1$ respectively, as a function of $\lambda_1$. Note that some eigenvalues are degenerate due to the symmetry of this example.

By investigating the spectrum of the $1$-multiplex Hodge Laplacian we observe that the  Fieder eigenvalue of the $1$-up Hodge Laplacian display a maximum for $\gamma_1=\gamma_2$ where there is a level crossing  of the two non-zero eigenvalues.
This phenomenon is illustrated in the synthetic  multiplex simplicial complexes  (see Figure $\ref{fig:3}$), where we plot the Fiedler eigenvalue $\lambda_F^{up}$ of the $1$-up Hodge Laplacian as a function of $\gamma_2-\gamma_1$ showing that in presence of simplices overlap the maximum Fiedler eigenvalue of observed for $\gamma_1=\gamma_2$, and the curve in non-monotonic. In the figure, we observe  that the removal of some simplices in one layer can have  different effects; increasing or decreasing the Fiedler eigenvalue while always decreasing the sharpness of the maximum.

In order to further investigate this property we have considered a random $2$-dimensional simplicial complex with tunable level of triangle overlap.
The random simplicial complex model is constructed  by drawing empty triangles with probability $p_2$ between each triple of nodes among a set of $\bar{N}$ nodes. The resulting clique complex forms the simplicial complex in the first layer. The second layer is constructed by considering each triangle present in the first layer and including it, together with all its faces in  the second layer  with probability $p_1$. Therefore this model can be used to explore the effect of triangle overlap (tuned with the parameter $p_1$) in the value of the Fiedler eigenvalue $\lambda_F^{up}$.
The results of this numerical investigation,  shown in Fig. $\ref{fig:3b}$, reveal that the maximum of $\lambda_F^{up}$ at $\gamma_1\simeq \gamma_2$ is a robust property observed also in this null model, provided that triangle overlap is large (large $p_1$). We present preliminary numerical results on a few simplicial complexes here, but a systematic study of overlap may be the subject of future research.

\subsection{Results on real multiplex datasets}
In this section, we consider the spectral properties of two different real multiplex simplicial complexes, and their effect on the higher-order diffusion of the topological signals.
We consider the multiplex microbiome dataset \cite{ding2014dynamics,de2016spectral} (available at \cite{manlio_github}) showing interactions present amongst different types of microbial communities in the human body. We build five multiplex simplical complexes (4 with $M=2$ layers-duplex simplicial complexes- and 1 with $M=3$) combining four distinct human body microbial communities (see Table \ref{table1a} for the layer composition).  We also consider multiplex simplicial complexes built from the multiplex C. elegans connectome \cite{chen2006wiring,de2015muxviz}, (available at \cite{manlio_github}) where the multiplex consists of layers corresponding to: electric (gap-junction) connection ("ElectrJ"), chemical monadic ("MonoSyn"), and polyadic ("PolySyn") synaptic connections. Specifically, out of the multilayer c.elegans connectome we build three multiplex simplicial complexes of $M=2$ layers (duplex simplicial complexes) and one multiplex simplicial complex of $M=3$ layers (see Table \ref{table1a} for the layer composition). All these multiplex simplicial complexes are built from the multiplex networks by considering the $d=2$ dimensional clique complex of each layer, i.e. filling all the cliques of the multiplex network structure. Interestingly, the distribution of multisimplices in these multiplex simplicial complexes is non-trivial, often showing a significant overlap of nodes, links and triangles of different type $\vec{m}$.
This can be observed from the Tables \ref{table1b} and \ref{table2} displaying  the total number of multinodes $N^{\vec{m}}$ the total number of multilinks $L^{\vec{m}}$ and the total number of multitriangles $T^{\vec{m}}$ for the considered multiplex simplicial complexes of $M=2$ and $M=3$ layers respectively.

Here, we perform a systematic study of the dependence of the irrotational and solenoidal Fiedler eigenvalues of the $n=1$ multiplex Hodge Laplacian on the coupling between the layers, for several real multiplex simplicial complexes. In our analysis, we chose $W_n=1$ for all value of $n$, however changing their values, for instance changing the ratio $W_2$ keeping $W_1$ and $W_0$ constant, will modulate the scale of the  solenoidal spectrum and hence can modulate  the value of the Fiedler eigenvalue $\lambda_F^+$ allowing this to  acquire larger or smaller values, as desirable in applications.

Our analyses  (see Figures $\ref{fig:4}-\ref{fig:7}$) reveal diverse functional behavior of the Fiedler eigenvalues with the coupling constant $\gamma_n$. Indeed, in some cases the multiplex  Fiedler eigenvalue can be larger than the uncoupled one obtained for $\gamma_n=0$ for any value of $n$, indicating that the proposed topological coupling between the layers can speed up the higher-order diffusion dynamics. However we also find instances of multiplex simplicial complexes in which the opposite behavior is observed and the coupling between the layers leads to a slower dynamics (smaller Fiedler eigenvalue that in the uncoupled scenario).
This is particularly interesting for the study of  real world complex networks where controlling diffusion of solenoidal or irrotational signals is desirable. For instance, in brain networks, the mixing of bottom-up sensory information and top-down predictions, i.e. prior beliefs are misinterpreted as sensory observations, also known as circular belief propagation, underlie psychotic behaviors in schizophrenia \cite{jardri2013circular,jardri2017experimental}, This can be interpreted as circular signals, and it may be desirable then to diffuse these to zero quickly. Thus, our framework offers one method for the study of, and control of, solenoidal and irrotational signals that may be important in several biological and other complex systems

\section{Conclusions}

Characterizing the topology of multiplex higher-order networks is  key in order to capture the interplay between higher-order structure and higher-order diffusion on multilayer data.
Here, we propose a framework that leverages on the ubiquitous presence of  the overlap of simplices,  in order to couple the higher-order diffusion dynamics occurring on the different layers of multiplex higher-order networks. Specifically,  we  introduce multiplex metric matrices that couple overlapping simplices. These metric matrices are used to  define multiplex Hodge Laplacians and Dirac operators, thus enforcing a higher-order multiplex diffusion dynamics across the layers of multiplex simplicial complexes. The multiplex Hodge Laplacians and Dirac operators  extend the standard definition of the monoplex Hodge Laplacians and Dirac operators, and have spectral properties encoding for the topology of the multiplex simplicial complexes.
Here we studied the properties of the resulting higher-order diffusion on   synthetic multiplex simplicial complexes and on clique complexes of real multiplex networks.
Specifically, we have  analysed the properties of  higher-order diffusion on a random model of multiplex simplicial complexes with tunable level of overlap among the simplices, and on real  connectome and microbiome multiplex simplicial complexes.  In summary, our results indicate that the properties of higher-order diffusion can be controlled by modulating the coupling among the layers and/or  by varying the topology of the multiplex simplicial complexes. In particular, our results reveal that the coupling among different layers can either speed up or slow down higher-order multiplex diffusion depending on the topology of the multiplex simplicial complex.

We believe that this work introduces a fundamental mathematical framework for treating higher-order diffusion on multiplex simplicial complexes and demonstrates applicability of this framework to a wide range of data-driven complex systems. This work opens new perspectives for uncovering the topology of multiplex higher-order simplicial complexes and it is our hope that the multiplex Hodge Laplacians and multiplex Dirac operators defined hereby could find further applications in the modelling and control of higher-order nonlinear dynamics on multiplex structures. 

\section*{Author contribution statement}
\noindent Sanjukta Krishnagopal: Methodology, Software, Investigation, Formal analysis, Writing. \\
Ginestra Bianconi: Conceptualization, Methodology, Software, Investigation, Formal analysis, Writing.

\section*{Declaration of competing interest}
The authors declare that they have no known competing financial interests or personal relationships that could have appeared to
influence the work reported in this paper.

\section*{Data availability}
The data is publicly available at the repository \cite{manlio_github}.

\appendix
\section{Algebraic topology of multiplex simplicial complexes}
\subsection{Multiplex simplicial complexes}
We define a multiplex simplicial complex  $\mathcal{K}$ formed by $M$ layers $\alpha\in \{1,2,\ldots,M\}$ as the set formed by the union  of the simplicial complexes forming each single layer of the multiplex structure. In other words,  ${\mathcal{K}}={K}^{[1]}\cup {K}^{[2]}\cup\ldots {K}^{[M]}$ where  ${K}^{[\alpha]}$ indicates the simplicial complex given by the set of  nodes, links, triangles, tetrahedra and higher-order simplices present in layer $\alpha$, closed under the inclusion of the faces of each simplex.
We indicate with $\sigma_i^n$ the generic $n$-dimensional simplex of the multiplex simplicial complex regardless of the layer $\alpha$ to which it belongs with  
 {\bea
\sigma_i^n=[v_0;\alpha,v_1;\alpha,\ldots, v_n;\alpha],
\eea
where here we adopt the same notation as in the main text for the generic replica node $[v_j;\alpha]$ in layer $\alpha$.}
To each simplex in layer $\alpha$ we associate an orientation induced by the labelling of the nodes in the layer, i.e.  {a simplex $\sigma_i^n$ has positive orientation for $v_0<v_1<\ldots v_n$ and for any even permutation of the labels of the nodes.} Note that replica nodes have identical labelling within their own layer, thus overlapping simplices will share the same orientation.\\
 {For ease of notation, in this Appendix, we will omit the labels of the layer for the replica nodes, and thus we will indicate the simplex $\sigma_i^n$ as 
\bea
\sigma_i^n=[v_0,v_1,\ldots, v_n],
\eea
where if $\sigma_i^n$ belongs to layer $\alpha$, i.e. $\sigma_i^n\in K^{[\alpha]}$ then all the nodes belonging to it also belong to layer $\alpha$, i.e. $v_0,v_1,\ldots v_n\in K^{[\alpha]}$ or in other words they are  replica nodes in layer $\alpha$. }
The set of all $n$-dimensional simplices of the multiplex simplicial complex is here indicated as $\mathcal{Q}_n$ while the set of $n$-dimensional simplices of layer $\alpha$ is indicated as ${Q}_n^{[\alpha]}$. Clearly, $\mathcal{Q}_n$ is the union of the $n$ simplices of each layer $\alpha$, i.e.
$\mathcal{Q}_n=\bigcup_{\alpha=1}^MQ_n^{[\alpha]}$.
We define the dimension $d$ of the multiplex simplicial complex $\mathcal{K}^M$, as the largest dimensions of the simplicial complexes describing each single layer, i.e. the largest dimension of any simplex in $\mathcal{K}$.

\subsection{Chains and cochains}
We define the multiplex $n$-chain as the  element of the free Abelian group $C_n$ with
basis the $n$-dimensional simplices of the multiplex simplicial complex defined with
respect to the ring $\mathbb{Z}$. 
It follows that a multiplex $n$-chain $c_n\in C_n$ can be expressed as 
\bea
c_n=\sum_{\sigma_i^n\in \mathcal{Q}_n}\sigma_i^n c_n^i,
\eea
with $c_n^i\in \mathbb{Z}$.
Note that a chain  $c_n\in C_n$ will in general  include linear combinations of simplices belonging to different layers and having non-zero coefficients.
In particular the group of multiplex $n$-chains $C_n$ is distinct from the group of $n$-chains in layer $\alpha$ indicated as $C_n^{[\alpha]}$ whose elements $c_{n,\alpha}\in C_n^{[\alpha]}$ can be expressed as 
\bea
c_{n,\alpha}=\sum_{\sigma_i^n\in {Q}_n^{[\alpha]}}\sigma_i^n c_{n,\alpha}^i,
\eea
with $c_{n,\alpha}^i\in \mathbb{Z}$.
The multiplex boundary operator $\partial_n:C_n\to C_{n-1}$ is a linear operator that maps multiplex $n$-chains to multiplex $(n-1)$-chains, which acts on the canonical basis of $n$-chains as
\bea
\partial_n[v_0, \ldots, v_n]=\sum_{p=0}^n (-1)^p[v_0,\ldots,\hat{v}_p,\ldots, v_n],
\eea
where $\sigma_i^n=[v_0, \ldots, v_n]\in Q_n$ and where $\hat{v}_p$ indicates the absence of node $v_p$.
Note that the $n$-th boundary operator of layer $\alpha$, indicated as $\partial_n^{[\alpha]}$ has the same definition as the multiplex boundary operator but its action is restricted to the $n$-chains of layer $\alpha$ only, i.e. $\partial_n^{[\alpha]}:C_n^{[\alpha]}\to C_{n-1}^{[\alpha]}$. Therefore we have 
\bea
\partial_n^{[\alpha]}[v_0, \ldots, v_n]=\sum_{p=0}^n (-1)^p[v_0,\ldots, \hat{v}_p,\ldots, v_n],
\eea
with $\sigma_i^n=[v_0, \ldots, v_n]\in Q_n^{[\alpha]}$.

Since each simplex in layer $\alpha$ admits as faces only simplices belonging to layer $\alpha$, it follows that the multiplex boundary operator is given by the direct sum of the boundary operators acting on simplicies of each distinct layer, i.e.
\bea
\partial_n=\bigoplus_{\alpha=1}^M \partial_n^{[\alpha]}.
\eea
We have therefore that the multiplex boundary operator obey for any $n\geq 1$:
\bea
\partial_n\partial_{n+1}=0,
\eea
i.e. the multiplex boundary of the multiplex boundary is null. This is a direct consequence of the definition of the multiplex boundary operator and the property 
\bea
\partial_n^{[\alpha]}\partial_{n+1}^{[\alpha]}=0,
\eea
valid for each individual layer $\alpha$ of the multiplex network.

A multiplex $n$-cochain ${f}\in C^n$  is a  homeomorphism between the multiplex $n$-chains $C_n$ and the reals $\mathbb{R}$, i.e. ${f}:C_n\to \mathbb{R}$ such that 
\bea
f(c_n)=\sum_{\sigma_i^n\in \mathcal{Q}_n}c_n^i f(\sigma_i^n).
\eea
Therefore, given a basis for the $n$-dimensional multiplex simplices, a multiplex $n$-cochain $f$ is uniquely determined by the vector ${\bf f}$  with elements given by  $f_i=f(\sigma_i^n)$.
A multiplex $n$-cochain should be distinguished from a $n$-cochain $f\in C^{n,[\alpha]}$ of each individual layer $\alpha$  of the multiplex network defined as $f:C_n^{[\alpha]}\to \mathbb{R}$ such that 
\bea
f(c_{n,\alpha})=\sum_{\sigma_i^n\in {Q}_n^{[\alpha]}}c_{n\alpha}^i f(\sigma_i^n).
\eea

The multiplex coboundary operator $\delta_n:C^n\to C^{n+1}$ is a linear operator mapping $n$-dimensional multiplex cochains to $n+1$ multiplex cochains whose action on the basis of multiplex cochains is determined by 
\bea
(\delta_n f)(\sigma_i^{n+1})= f\circ\partial_n (\sigma_i^{n+1}),
\eea
where $\sigma_{n+1}^i=[v_0,v_1,\ldots, v_{n+1}]\in \mathcal{Q}$.
 The above definition  can be equivalently expressed as 
\bea
(\delta_n f)[v_0,v_1,\ldots, v_{n+1}]= \sum_{p=0}^n(-1)^n f([v_0,v_1,\ldots \hat{v}_p\ldots, v_{n+1}]).
\label{M_coboundary}
\eea

The coboundary operator $\delta_n^{[\alpha]}:C^{n,[\alpha]}\to C^{n+1,[\alpha]}$ is defined analogously by its action on the $(n+1)$-dimensional simplices of layer $\alpha$ by 
\bea
(\delta_n f)[v_0,v_1,\ldots, v_{n+1}]= \sum_{p=0}^n(-1)^n f([v_0,v_1,\ldots \hat{v}_p\ldots, v_{n+1}]),
\label{M_coboundary_alpha}
\eea
where $\sigma_{n+1}^i=[v_0,v_1,\ldots, v_{n+1}]\in Q^{[\alpha]}$.
The multiplex coboundary operator defined in Eq.(\ref{M_coboundary}) is the direct sum of the coboundary operators acting on each single layer, i.e. we  have 
\bea
\delta_n=\bigoplus_{\alpha=1}^M \delta_n^{[\alpha]},
\eea
where $\delta_n^{[\alpha]}$  indicates the coboundary operator in layer $\alpha$.
The multiplex coboundary operator defined in this way obeys 
\bea
\delta_{n+1}\delta_{n}=0.
\eea
The multiplex coboundary operator $\delta_n$ can be expressed in matrix form by the $(n+1)$ coboundary matrix 
 ${\mathcal {B}}_{n+1}$ is the block diagonal matrix given by 
\bea
{\mathcal {B}}_{n+1}=\left(\begin{aligned}& {B}^{[1]}_{n+1} && 0 && 0 && 0\\& 0 && {B}^{[2]}_{n+1} &&0 &&0 \\  & 0 && 0  && \ldots &&0 \\& 0 && 0  && 0 && {B}^{[M]}_{n+1} 
\end{aligned}\right),
\eea
where ${B}^{[\alpha]}_{n+1}$ is the $(n+1)$-th coboundary matrix matrix for layer $\alpha$.
For instance, for a  2-layer multiplex simplicial complex,  the  $(n+1)$-{th} coboundary matrix is the  matrix given by 
\bea
{\mathcal {B}}_{n+1}=\left(\begin{aligned}& {B}^{[1]}_{n+1} && 0\\& 0 && {B}^{[2]}_{n+1}  \end{aligned}\right).
\eea
\subsection{Adjoint coboundary operator}
As discussed the the main text of this work, the coupling between the different layers of the multiplex simplicial complex is implemented by designing appropriate multiplex metric matrices.
The metric matrices ${\mathcal{G}}_n^{-1}$ define non-degenerate scalar products between $n$-dimensional multiplex cochains.
In particular, given $f_1,f_2\in C^{n}$ their scalar product is defined as 
\bea
\Avg{f_2,f_1}={\bf f}_2^{\top}\mathcal{G}_{n}^{-1}{\bf f}_1,
\eea
where $\mathcal{G}_n$ are positive definite and non-singular matrices obeying $\mathcal{G}_n^{\top}=\mathcal{G}_n$.
In particular the matrices $\mathcal{G}_n$ are coupling the layers when simplices in different layer overlap (see for a detail explanation of their formulation Section $\ref{sec:metric}$)
The multiplex adjoint coboundary operator $\delta^{\star}_n:C^{n+1}\to C^n$ is defined as the operator for which 
\bea
\Avg{g,\delta_n f}=\Avg{\delta^\star_n g ,f}
\label{uno}
\eea
holds for every $f\in C^n$ and $g\in C^{n+1}$.
From this definition it follows that 
\bea
\delta_{n}^{\star}\delta_{n+1}^{\star}=0,
\eea
for $n\geq 0$.
Indeed, using the definition of the adjoint coboundary operator repeatedly we have that,  for every $f\in C^{n}$ and $g\in C^{n+2}$ 
\bea
\Avg{\delta^\star_{n}\delta^\star_{n+1} g,f}=\Avg{g,\delta_{n+1}\delta_{n} f}=0,
\eea
where in the last expression we use $\delta_{n+1}\delta_n=0$.
In matrix form, the multiplex adjoint coboundary operator $\delta_n^{\star}$ is expressed by the matrix $\mathcal{B}_{n+1}^{\star}$ which depends on the multiplex coboundary matrix $\mathcal{B}_{n+1}$ and by the metric matrices $\mathcal{G}_{n+1}$ and $\mathcal{G}_{n}$ with
\bea
\mathcal{B}_{n+1}^{\star}=\mathcal{G}_{n}\mathcal{B}_{n+1}^{\top}\mathcal{G}_{n+1}^{-1}.
\label{Bstar}
\eea
This expression follows immediately from the  definition of the multiplex adjoint coboundary operator as from Eq. (\ref{uno}) we get 
\bea
{\bf g}\mathcal{G}_{n+1}^{-1}\mathcal{B}_{n+1}{\bf f}={\bf f}(\mathcal{B}_{n+1}^{\star})^{\top}\mathcal{G}_n^{-1}{\bf g}.
\eea
Since this equation is valid for every ${\bf f}$ and ${\bf g}$ we have then 
\bea
\mathcal{G}_{n+1}^{-1}\mathcal{B}_{n+1}=(\mathcal{B}_{n+1}^{\star})^{\top}\mathcal{G}_n^{-1}.
\eea
From which it follows 
\bea
\mathcal{B}_{n+1}^{\star}=\mathcal{G}_{n}\mathcal{B}_{n+1}^{\top}\mathcal{G}_{n+1}^{-1},
\eea
which is equivalent to Eq.(\ref{Bstar}).
\subsection{Multiplex Hodge Laplacians }
Having defined the multiplex coboundary operator and its adjoint operator, we are now ready to define the multiplex Hodge Laplacian $L_n$ that the extends the notion of the Hodge Laplacian of a  monoplex simplicial complex~\cite{eckmann1944harmonische,horak2013spectra,jost2019hypergraph,lim2020hodge,meng2020weighted}. In particular we  define the multiplex Hodge Laplacian as 
\bea
L_n=L_n^{up}+L_n^{down},
\eea
where 
\bea
L_n^{up}=\delta_n^{\star}\delta_n, \quad L_n^{down}=\delta_{n-1}\delta_{n-1}^{\star}.
\eea
Both $L_n^{up}$ and $L_n^{down}$ are positive semi-definite self-adjoint operators, from which it follows that also the multiplex Hodge Laplacian $L_n$ is positive semi-definite.
From the properties of the multiplex coboundary operator and its adjoint operator  it follows immediately that the multiplex Hodge Laplacian obeys Hodge decomposition, i.e.
\bea
\mbox{ker}(L_n^{up})&\supseteq &\mbox{im}(L_n^{down}),\nonumber \\
\mbox{ker}(L_n^{down})&\supseteq &\mbox{im}(L_n^{up}),\nonumber \\
\mbox{ker}({L}_n)&=&\mbox{ker}({L}_n^{up})\cap\mbox{ker}({L}_n^{down}).
\label{Hodge}
\eea
Indeed the first two relations are derived by noticing that
\bea
L_n^{up}L_n^{down}&=&\delta_n^{\star}\delta_n\delta_{n-1}\delta_{n-1}^{\star}=0,\nonumber \\
L_n^{down}L_n^{up}&=&\delta_{n-1}\delta_{n-1}^{\star}\delta_n^{\star}\delta_n=0,
\eea
where the last equality directly follow from the identities $\delta_n\delta_{n-1}=0$ and $\delta_{n-1}^{\star}\delta_n^{\star}=0$.
The third relation of Eqs. (\ref{Hodge}) is a direct consequence of the fact that $L_n=L_n^{up}+L_n^{down}$ where both $L_n^{up}$ and $L_n^{down}$ are semidefinite positive.
An important consequence of Hodge decomposition is that every $n$-cochain $f\in C^n$ can be decomposed in a unique way into an irrotational flow $f_1\in \mbox{im}(\delta_{n-1})$ an harmonic flow $f_{harm}\in \mbox{ker}(L_n)$, a solenoidal flow $f_2\in \mbox{im}(\delta_{n}^{\star})$, i.e.
\bea
f=f_1+f_2+f_{harm}.
\eea
This decomposition of $n$-dimensional cochain follows from the spectral properties of the multiplex Hodge Laplacians and implies
\bea
C^{n}=\mbox{im}(\delta_{n-1})\oplus \mbox{ker}(L_n)\oplus \mbox{im}(\delta_{n}^{\star}).
\eea

In matrix form the multiplex Hodge Laplacians can be expressed as
\bea
\mathcal{L}_{n}=\mathcal{L}_{n}^{up}+\mathcal{L}_{n}^{down}
\eea
with 
\bea
\mathcal{L}_{n}^{up}&=&\mathcal{G}_n\mathcal{B}_{n+1}^{\top}\mathcal{G}_{n+1}^{-1}\mathcal{B}_{n+1},
\nonumber \\
\mathcal{L}_{n}^{down}&=&\mathcal{B}_{n}\mathcal{G}_{n-1}\mathcal{B}_{n}^{\top}\mathcal{G}_{n}^{-1}.
\label{Laplacian_up_down}
\eea
It can be shown that  $\mathcal{L}_n^{up}$ is isospectral with the symmetrized Hodge Laplacian $\hat{\mathcal{L}}_n^{up}$  and that $\mathcal{L}_n^{down}$ is isospectral with the symmetrized Hodge Laplacian $\hat{\mathcal{L}}_n^{down}$ with
\bea
\hat{\mathcal{L}}_n^{up}&=&\mathcal{G}_n^{1/2}\mathcal{B}_{n+1}^{\top}\mathcal{G}_{n+1}^{-1}\mathcal{B}_{n+1}\mathcal{G}_n^{1/2},\nonumber \\
\hat{\mathcal{L}}_n^{down}&=&\mathcal{G}_{n}^{-1/2}\mathcal{B}_{n}\mathcal{G}_{n-1}\mathcal{B}_{n}^{\top}\mathcal{G}_{n}^{-1/2}.
\eea Hence both the up and down multiplex Hodge-Laplacians have real (non negative) eigenvalues.

\subsection{Multiplex Dirac Operators}
Recently, growing attention has been devoted to the study of discrete Dirac operators on monoplex simplicial complexes~\cite{bianconi2021topological,baccini2022weighted,wee2023persistent}.
We now define the multiplex Dirac operators $D_n:C_{n-1}\oplus C_{n}\to C_{n-1}\oplus C_{n}$ acting on topological spinors $\psi=f\oplus g$ with $f\in C_{n-1}$ and $g\in C_{n}$ as 
\bea
D_n\psi=D_n (f\oplus g)=\delta_{n-1}^{\star}g\oplus \delta_{n-1}f.
\eea
From this definition it is observed that the multiplex Dirac operators allow cross-talk between topological signals of different dimension and can thus have wide use in the study of coupled multiplex topological signals of different dimensions modeled as cochains of dimension $n-1$ and dimension $n$. On monoplex simplicial complexes the Dirac operator has already been used extensively to study synchronization, Turing patterns and signal processing, in addition to providing a way to formulate a discrete topological field theory \cite{bianconi2023dirac,bianconi2023mass,calmon2022dirac,giambagli2022diffusion,muolo2023three,calmon2023dirac}. 
Having defined the multiplex Dirac operators $D_n$ coupling cochains of dimension $n-1$ and dimension $n$ we can now define the  multiplex Dirac operator $D$ as the direct sum of the multiplex Dirac operator of any arbitrary dimension, i.e.
\bea
D=\sum_{n=1}^d D_n.
\eea

One fundamental property of the multiplex Dirac operator $D_n$ is that
their square is given by the direct sum of the Hodge Laplacians, in particular we have 
\bea
D_n^2=L_{n-1}^{up}\oplus L_n^{down}.
\label{Dsquare}
\eea
Indeed using the definition of the multiplex Dirac operator, for every $\psi=f\oplus g$ with $f\in C_{n-1}$ and $g\in C_{n}$ we have
\bea
D_n^2\psi=D_n^2 (f\oplus g)=D_n \delta_{n-1}^{\star}g\oplus \delta_{n-1}f=\delta_{n-1}^{\star}\delta_{n-1}f \oplus \delta_{n-1}\delta_{n-1}^{\star}g=L_{n-1}^{up}f\oplus L_n^{down}g.
\eea
Since it can be shown that $L_{n-1}^{up}$ and $L_n^{down}$ are isomorphic, from Eq. (\ref{Dsquare}) it follows that the non-zero eigenvalues $\eta_n$ of the multiplex Dirac operator $D_n$ are given by 
\bea
\eta_n=\pm\sqrt{\lambda_n},
\label{eig_D}
\eea
where $\lambda_n$ indicates the generic non-zero eigenvalues of $L_n^{down}$ that, as we will see in the following are real and positive. Hence for every non-zero eigenvalue $\lambda_n$ of $L_n^{down}$ the multiplex Dirac operator admits one positive and one negative eigenvalue given by Eq.(\ref{eig_D}).
Given that  the multiplex Dirac operators $D_n$ obey Eq. $(\ref{Dsquare})$, the square of the  multiplex Dirac operator is given by 
\bea
D^2=\bigoplus_{n=1}^d L_n.
\eea
Therefore $D$ can be interpreted as the ``square root" of the super Hodge Laplacian $\tilde{L}=\bigoplus_{n=1}^d L_n$.
Consequently, the non-zero eigenvalues of the multiplex Dirac operators are given by 
\bea
\eta=\pm\sqrt{\lambda},
\eea
where $\lambda$ is the generic eigenvalue of any Hodge Laplacian $L_n$ with $0\leq n\leq d$.
The multiplex Dirac operators $D_n$ can be expressed in matrix form as 
\bea
\mathcal{D}_n=\left(\begin{array}{cc}0& \mathcal{G}_{n-1}\mathcal{B}_{n}^{\top}\mathcal{G}_n^{-1}\\
	{\mathcal{B}_{n}}& 0\end{array}\right).
\eea
On a multiplex simplicial complex of dimension $d=2$, the Dirac operator reads in matrix form as 
\bea
\mathcal{D}=\left(\begin{array}{ccc}0& \mathcal{G}_{0}\mathcal{B}_{1}^{\top}\mathcal{G}_1^{-1} &0\\
	{\mathcal{B}_{1}}& 0 &\mathcal{G}_{1}\mathcal{B}_{2}^{\top}\mathcal{G}_2^{-1}\\
 0& {\mathcal{B}_{2}} &0
 \end{array}\right).
\eea
\subsection{Multiplex Betti numbers}
In this section our goal is to evaluate the Betti number of the multiplex simplicial complex.
In order to proceed with this derivation, let us recall that the co-homology groups $\tilde{H}^{\alpha}(K^{[\alpha]},\mathbb{R})$  of the simplicial complex in layer $\alpha$ are defined as 
\bea
\tilde{H}_n({K}^{[\alpha]},\mathbb{R}):=\frac{\mbox{ker} (\delta_n^{[\alpha]})}{\mbox{im}(\delta_{n-1}^{[\alpha]})}. 
\eea
The Betti number $\beta_n$ of the layer $\alpha$ gives the number of $n$-dimensional cavities in the considered layer and is given by 
\bea
\beta_n^{[\alpha]}=\mbox{dim} \ \tilde{H}_n({K}^{[\alpha]},\mathbb{R}).
\eea
Similarly here we define the multiplex co-homology groups of the multiplex simplicial complex as 
\bea
\tilde{\mathcal{H}}_n(\mathcal{K},\mathbb{R}):=\frac{\mbox{ker} (\delta_n)}{\mbox{im}(\delta_{n-1})}
\eea
and the multiplex Betti number $\beta_n^M$ as 
\bea
\beta_n^{M}=\mbox{dim} \ \tilde{\mathcal{H}}_n({K},\mathbb{R}).
\eea
Since the multiplex coboundary operator is the direct sum of the coboundary operators of each individual layer $\alpha$, it follows that the multiplex Betti number are the sum of the corresponding Betti numbers of the individual layers, i.e.
\bea
\beta_n^M=\sum_{\alpha=1}^M \beta_n^{[\alpha]}.
\eea

Let us now investigate the relationship between the multiplex Betti numbers and the spectral properties of the multiplex Hodge Laplacian.

Since $L_n^{up}=\delta_n^{\star}\delta_n$ and $L_n^{down}=\delta_{n-1}\delta_{n-1}^{\star}$ are self-adjont operators, any $n$-cochain $f\in \mbox{ker}(L_n)=\mbox{ker}(L_n^{up})\cap \mbox{ker}(L_n^{up})$ obeys
\bea
\Avg{f,L_n f}=\Avg{f,L_n^{up}f}+\Avg{f,L_n^{down}f}=0,
\eea
with 
\bea
\Avg{f,L_n^{up}f}=0,\quad \Avg{f,L_n^{down}f}=0.
\eea
Using the definition of the adjoint multiplex coboundary operator (Eq.(\ref{uno})), these two last equations imply that  
\bea
\Avg{f,L_n^{up}f}=\Avg{f,\delta_n^{\star}\delta_n f}=\Avg{\delta_n f,\delta_n f}=0,\nonumber \\
\Avg{f,L_n^{down}f}=\Avg{f,\delta_{n-1}\delta_{n-1}^{\star} f}=\Avg{\delta_{n-1}^{\star} f,\delta_{n-1}^{\star} f}=0,
\eea
and hence $f\in \mbox{ker}(\delta_n)$ and $f\in \mbox{ker}(\delta_{n-1}^{\star})$ or in other words,
\bea
\mbox{ker}(L_n)=\mbox{ker}(\delta_n)\cap \mbox{ker}(\delta_{n-1}^{\star}).\label{1}
\eea
Moreover, it is easy to show that 
\bea 
\mbox{ker}(\delta_{n-1}^{\star})=(\mbox{im} (\delta_{n-1}))^{\perp}.
\label{2}
\eea
In fact, if $f\in \mbox{ker}(\delta_{n-1}^{\star})$ we have that 
\bea
\Avg{g,\delta_{n-1}^{\star} f}=0,
\eea
for every choice of $g\in C^{n}$. Using the definition of the adjoint multiplex coboundary operator given by Eq.(\ref{uno}) we obtain
\bea
\Avg{g,\delta_{n-1}^{\star} f}=\Avg{\delta_{n-1}g,f}=0,
\eea
from which we get $f\in (\mbox{im} (\delta_{n-1}))^{\perp}$.
Using Eq.(\ref{1}) and Eq. (\ref{2}) we can then express the kernel of the multiplex Hodge Laplacian as 
\bea
\mbox{ker}(L_n)=\mbox{ker}(\delta_n)\cap (\mbox{im}( \delta_{n-1}))^{\perp}\simeq \tilde{\mathcal{H}}_n(\mathcal{K},\mathbb{R})
\eea
It follows that the dimension of the kernel of the multiplex Hodge Lapalcian is equal to the multiplex Betti numbers
\bea
\mbox{dim}\ \mbox{ker} (L_n)=\beta_n^M.
\eea
This result is valid in presence of an  arbitrary coupling between the layers induced by non trivial metric matrices $\mathcal{G}_n$ defined by the $\gamma_n$ parameters taking any allowed values $\gamma_n\in [0,1).$ Note that this result will remain valid also for other choices of the metric matrices $\mathcal{G}_n$ as long as these matrices are positive definite and non-singular.


\end{document}